\documentclass[%
superscriptaddress,
 amsmath,amssymb,
prx,
]{revtex4-2}

\usepackage{graphicx}% Include figure files
\usepackage{dcolumn}% Align table columns on decimal point
\usepackage{bm}% bold math
\usepackage{hyperref}
\hypersetup{
    colorlinks=true,
    linkcolor=blue,
    filecolor=magenta,      
    urlcolor=cyan,
    }
\usepackage[bottom]{footmisc}

\usepackage{xcolor}
\usepackage{dcolumn}% Align table columns on decimal point
\usepackage{bm}% bold math
\usepackage[separate-uncertainty=true]{siunitx}
\DeclareSIUnit \belm {Bm}
\usepackage[utf8]{inputenc}
\usepackage[T1]{fontenc}
\usepackage{mathptmx}
\usepackage{xspace}
\usepackage{amsmath}

\usepackage[final]{changes}
\definechangesauthor[color=blue]{common}
\definechangesauthor[color=red]{PVS}
\definechangesauthor[color=blue]{PH}
\definechangesauthor[color=purple]{KB}
\definechangesauthor[color=pink]{js}
\newcommand{\footnotePVS}[1]{\footnote{#1}}

\newcommand{\lSAW}{{\lambda_{\mathrm{SAW}}}}		% acoustic wavelength
\newcommand{\kSAW}{k_\mathrm{SAW}}			% acoustic wave vector
\newcommand{\vSAW}{v_\mathrm{SAW}}			% SAW velocity
\newcommand{\wSAW}{\omega_\mathrm{SAW}}			% acoustic angular frequency
\newcommand{\FSAW}{\Phi_\mathrm{SAW}}	 		% piezoelectric potential
\newcommand{\PSAW}{P_\mathrm{SAW}}			% acoustic power
\newcommand{\pSAW}{\phi_\mathrm{SAW}}			% SAW phase

\newcommand{\nanometer}[1]{\SI{#1}{\nano\metre}}
\newcommand{\micron}[1]{\SI{#1}{\micro\metre}}
\newcommand{\ev}[1]{\SI{#1}{\electronvolt}}
\newcommand{\mev}[1]{\SI{#1}{\milli\electronvolt}}

\newcommand{\commenta}[1]{}

\newcommand{\xdirnox}{$[110]$ }

\newcommand{\xydirnoxy}{$[010]$}

\newcommand{\BSO}{\vec{B}_\mathrm{SO}}
\newcommand{\OSO}{{  \Omega_\mathrm{SO}}}
\newcommand{\LSO}{L_\mathrm{SO}}

\begin{document}

\title{     Flying electron spin control gates  }

\author{Paul L. J. Helgers} 
\affiliation{Paul-Drude-Institut f{\"u}r Festk{\"o}rperelektronik, Leibniz-Institut im Forschungsverbund Berlin e.V., Hausvogteiplatz 5-7, 10117 Berlin, Germany }
\affiliation{NTT Basic Research Laboratories, NTT Corporation, 3-1 Morinosato-Wakamiya, Atsugi, Kanagawa 243-0198, Japan}

\author{James A. H. Stotz} 
\email{jstotz@queensu.ca}
\affiliation{Paul-Drude-Institut f{\"u}r Festk{\"o}rperelektronik, Leibniz-Institut im Forschungsverbund Berlin e.V., Hausvogteiplatz 5-7, 10117 Berlin, Germany }
\affiliation{Department of Physics, Engineering Physics \& Astronomy, Queen’s University, Kingston, ON, K7L3N6 Canada}

\author{Haruki Sanada}
\author{Yoji Kunihashi}
\affiliation{NTT Basic Research Laboratories, NTT Corporation, 3-1 Morinosato-Wakamiya, Atsugi, Kanagawa 243-0198, Japan}

\author{Klaus Biermann}
\author{Paulo V. Santos}
\email{santos@pdi-berlin.de}
\affiliation{Paul-Drude-Institut f{\"u}r Festk{\"o}rperelektronik, Leibniz-Institut im Forschungsverbund Berlin e.V., Hausvogteiplatz 5-7, 10117 Berlin, Germany }

\date{\today}% It is always \today, today,

\begin{abstract}
	\label{sec:abstract}
The control of "flying" (or moving) spin qubits is an important functionality for the manipulation and exchange of quantum information between remote locations on a chip. Typically, gates based on  electric or magnetic fields provide the necessary perturbation for their control either globally or at well-defined locations.  Here, we demonstrate the dynamic control of moving electron spins via contactless gates that move together with the spin. The concept is realized using  electron spins trapped and transported by moving potential dots defined by a surface acoustic wave (SAW). The SAW strain at the electron trapping site, which is set by the SAW amplitude, acts as a contactless, tunable gate that controls the precession frequency of the flying spins via the spin-orbit interaction. We show that the degree of precession control in moving dots exceeds previously reported results for unconstrained transport by an order of magnitude and is well accounted for by a theoretical model for the strain contribution to the spin-orbit interaction. This flying spin gate permits the realization of an acoustically driven optical polarization modulator based on electron spin transport, a key element for on-chip spin information processing with a photonic interface.
\end{abstract}

\maketitle
%\tableofcontents

\section{Introduction}

The spin field-effect transistor proposed by \textit{Datta and Das}~\cite{Datta90a} relies on the precession of moving (or flying) electron spins around the effective magnetic field $\BSO({\bf k})$ associated with the spin-orbit (SO) interaction, which depends on electron momentum ${\hbar\bf k}$. $\BSO$ can be electrically controlled by an electrostatic gate via the Bychkov-Rashba effect,\cite{Bychkov1984} thus opening the way for dynamic spin manipulation by electric fields. Spin transistors based on the electrical spin control have  so far been demonstrated only for ballistic spin transport along short ($<2~\mu$m) channels~\cite{Koo_S325_1515_09,Chuang_NN_10_35_15}. 

%%%%%%%%%%%%%%%%%%%%%%%%%%%%%%%
% Figure 0: Flying spin gates
\begin{figure*}[tbhp]
    \includegraphics[width=0.95\textwidth, angle=0, clip]{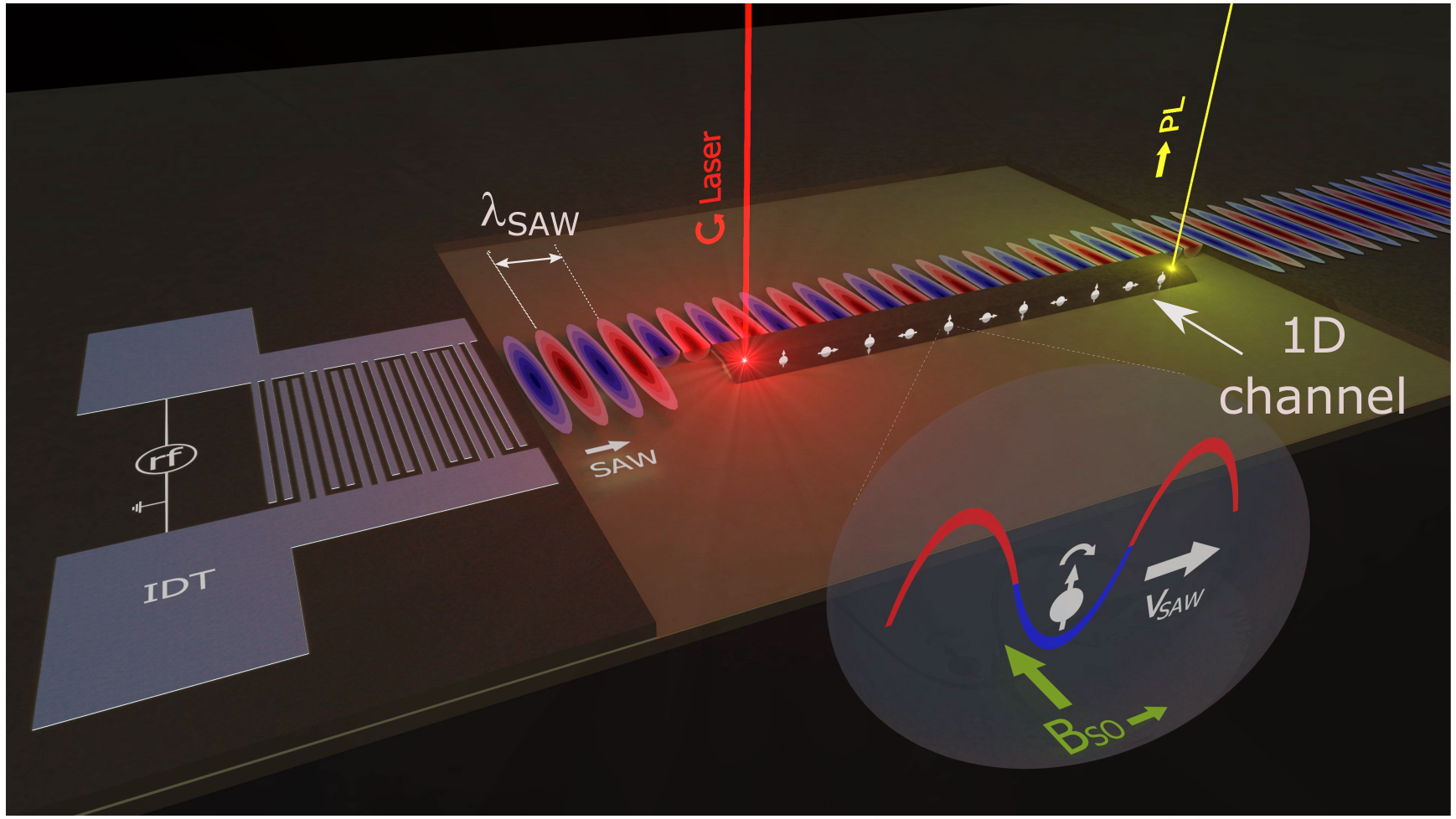}
    \caption{
        {\bf Flying control gate for electron spins.}
    A surface acoustic wave (SAW) excited by an interdigital acoustic transducer (IDT) is applied along a quasi-one dimensional (1D) semiconductor channel. The  piezoelectric potential of the SAW creates a moving quasi-zero dimensional (0D) potential dot, which captures spin-polarized, photoexcited electrons and holes  and transports them along the channel with the SAW velocity. Simultaneously, the SAW strain field induces an effective spin-orbit magnetic field $B_{SO}$ at the carrier transport sites with amplitude proportional to the SAW field. The latter acts as a flying spin gate to control the spin precession rate. 
    }
        \label{fig:Fig0}
    \end{figure*}
    %%%%%%%%%%%%%%%%%%%%%%%%%%%%%%%
    
SO-based spin control over long transport distances, which normally takes place in the diffusive regime, faces two main challenges.  The first is  D'yakonov-Perel' (DP) spin dephasing\cite{Dyakonov_SPJETP_33_1053} associated with the momentum-dependence of $\BSO$. Approaches to reduce DP spin dephasing and enable long-range spin transport lengths ($\ell_s$) include the engineering the SO interaction ~\cite{Ohno_PRL83_4196_99,PVS261,Bernevig_PRL97_236601_06,Koralek_N458_610_09,Walser2012,PVS261,Balocchi_PRL107_136604_11} as well as exploitation of motional narrowing effects~\cite{Dyakonov_SPJETP_33_1053}. The latter takes advantage of the inverse dependence of the DP dephasing rate on the carrier scattering time, which can be achieved via increased momentum scattering by impurities~\cite{JKDA99a} or at the boundaries of narrow transport channels  (i.e., channel widths less than the precession period, $\LSO$ under $\BSO$).
This latter approach has been realized using quantum wire channels ~\cite{Kiselev_PRB61_13115_00,Holleitner_PRL97_036805_06} as well as by enclosing the spins within moving potential dots \cite{PVS152,PVS265}.

The second major challenge for spin control is to devise  field configurations to simultaneously drive spin motion and generate a tunable $\BSO$ for controlled spin precession. 
An elegant solution is offered by the carrier transport by moving potential dots produced by a surface acoustic wave (SAW) along a one-dimensional (1D) channel. Figure~\ref{fig:Fig0} depicts an example based on a quantum wire (QWR) transport channel. Here, the moving piezoelectric potential modulation produced by the SAW stores photo-excited electrons and holes at different SAW subcycles and transports them with the acoustic velocity.  The spatial separation of electrons and holes prevents  recombination~\cite{Rocke98a} and, simultaneously, also suppresses spin relaxation due to the electron-hole exchange interaction~\cite{MSS93a}. The ambipolar SAW transport can thus transfer electron spins over long distances (up to \micron{100}) while enabling  optical spin readout  by detecting the polarization of photons emitted by the recombination of the transported carriers~\cite{PVS110,PVS272,PVS240,Bertrand_NN11_672_16,PVS265}.

%%%%%%%%%%%%%%%%%%%%%%%%%%%%%%%
% Figure 1: Schematic drawing of the principle
\begin{figure*}[tbhp]
\includegraphics[width=0.75\textwidth, angle=0, clip]{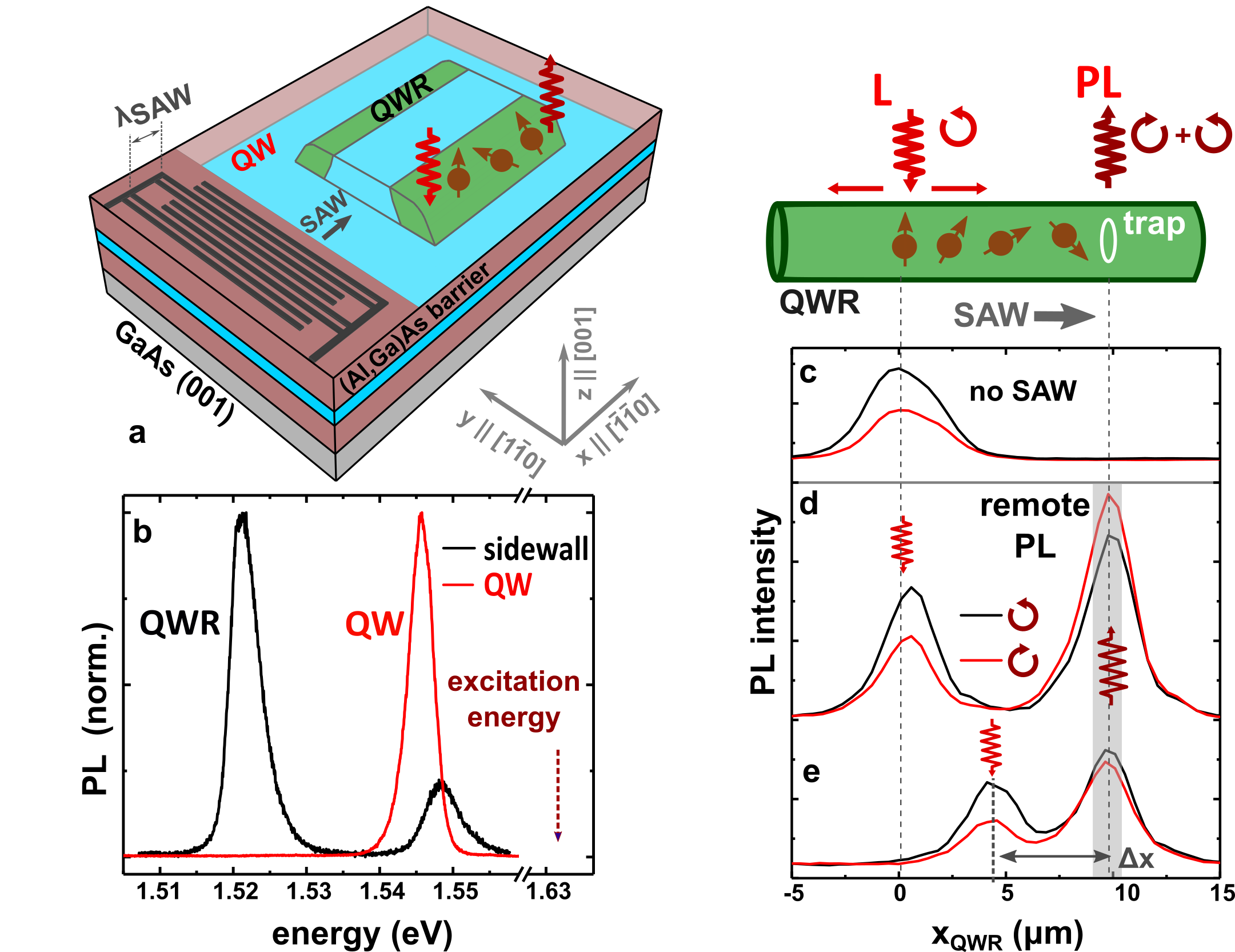}
\caption{
	{\bf Optically detected transport of spins in planar quantum wires.}
(a) Sidewall quantum wires (QWRs) formed by the epitaxial overgrowth of a quantum well (QW) on a GaAs (001) substrate structured with shallow ridges~\cite{PVS324}. Photoexcited carriers are transported along the QWR by surface acoustic waves (SAWs)  generated by interdigital acoustic transducers (IDTs). 
(b) Photoluminescence (PL) spectra recorded outside (red) and on (black) a ridge sidewall showing the emission lines of the QW (\ev{1.548}) and QWR (\ev{1.521}), respectively.
(c)-(e) Profiles of the right (black) and left (red) circularly polarized PL along the QWR axis ($x_\mathrm{QWR}$ coordinate) recorded under the configuration illustrated in the upper panel (c) in the absence and (d)-(e) under a SAW. The PL was excited by a right-circularly polarized laser spot focused at $x_\mathrm{QWR}=0$  for (c) and (d), and at $x_\mathrm{QWR}=4.5~\mu$m for (e). 
}
	\label{fig:Fig1}
\end{figure*}
%%%%%%%%%%%%%%%%%%%%%%%%%%%%%%%

Concomitantly with the transport, the SAW 
strain~\cite{Dyakonov_SPJETP_49_160,Pikus88a,Kato03a,Beck05a,English_PRB84_155323_11} 
and piezoelectric fields at the carrier location induce a  $\BSO$ contribution, which moves congruently with the carriers [cf. Fig.~\ref{fig:Fig0}].  The SAW amplitude acts, therefore,  as a contactless, flying spin gate, which  dynamically controls the rate of spin precession during transport. 
In this work, we demonstrate that long-range acoustic transport can be combined with a high degree of dynamical control of the spin precession rate (by over 250\%) if the flying spins are transported while confined within micron-sized moving potential dots.  
This degree of precession control exceeds by over an order of magnitude previous results for acoustic transport in 2D quantum well (QW) channels~\cite{PVS199,PVS240}.
We also show that the precession rate is mainly mediated by the acoustic strain imparted by the SAW in the storage phase of the electron spins during transport. The latter essentially enables the carrier wave to also act as a flying spin gate controlled by the SAW amplitude. 

The study addresses two types of acoustically defined, moving potential dots. In one system, the dots are formed by propagating a SAW along a quasi-planar GaAs quantum wire (QWR) [as illustrated in Fig.~\ref{fig:Fig0}]. 
We have also investigated spin transport and manipulation in moving potential dots created by the interfering piezoelectric fields of orthogonal SAW beams (denoted as dynamic quantum dots, DQDs\cite{PVS158,PVS152}). 
For both types of moving dots, we experimentally demonstrate the flying spin gates using optically excited spins that can be acoustically transported over large distances (tens of microns) with the spin precession rate controlled over a wide dynamic range dictated by the SAW amplitude (Sec.~\ref{Results}) and, notably, without external electric or magnetic fields. The measured precession rates are well accounted for by an analytical model for the SO fields generated by the SAW strain and piezoelectric fields, from which the strain-related SO parameters can be experimentally determined (Sec.~\ref{sofields}). As a further check of consistency, we show that spin precession rates and their dependence on the acoustic fields are also in good agreement with microscopic calculations of the spin splittings under the SAW field using a tight-binding approach. Interestingly, while the moving dot geometry enables precession control, a substantial enhancement of the spin lifetime due to motional narrowing is only observed for the DQDs. The limited spin lifetimes in the QWRs, in contrast, is attributed to the fact that the positive impact of lateral confinement  on the spin lifetime is offset by  spin scattering at the lateral, compositional interfaces. Even so, the dramatic ability of the flying spin gates to control the precession frequency by the SAW amplitude demonstrates a processor for optically encoded polarization information based on the dynamic control of electron spins during acoustic transport.

%%%%%%%%%%%%%%%%%%%%%%%%%%%%%%%
% Figure 2: spin polarization data of QW
\begin{figure*}[!tbhp]
	\includegraphics[width=.9\textwidth, angle=0, clip]{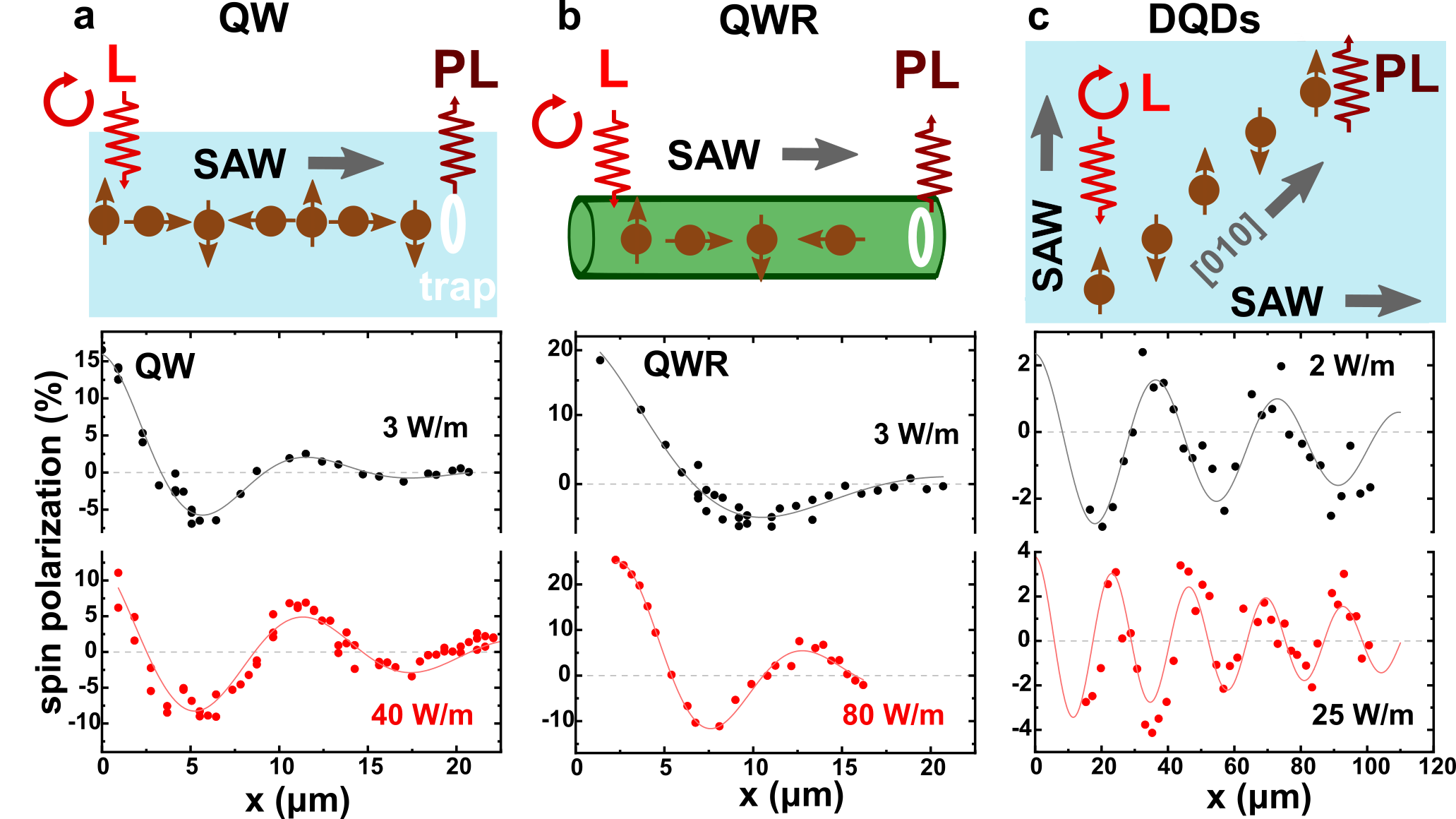}
	\caption{
		{\bf Acoustic control of the precession rate of moving spins.}
		Spin polarization,  $\rho_s$, during transport along (a) a quantum well (QW, excitation wavelength  $\lambda_L =  776$~nm, $P_{exc}$ = \SI{15}{\micro\watt}, 20~K), (b) a quantum wire (QWR, $\lambda_L =  776$~nm, $P_{exc}$ = \SI{30}{\micro\watt} (\SI{3}{\watt\per\metre}) and $P_{exc}$ = \SI{150}{\micro\watt} (\SI{80}{\watt\per\metre}), 20~K), and (c) dynamic dots (DQDs,  $\lambda_L = 776$~nm, 12~K) . The upper panels  illustrate the experimental setup:  spin-polarized electrons (brown) are optically injected by a circularly polarized laser ($L$, red) an transported by a SAW along the \xdirnox direction (QW and QWR) or along the \xydirnoxy\xspace direction (dynamic dots). The curves correspond to different SAW powers $\PSAW$. }
	\label{fig:Fig2}
\end{figure*}
%%%%%%%%%%%%%%%%%%%%%%%%%%%%%%% 

%%%%%%%%%%%%%%%%%%%%%%%%%%%%%%%%%%%%%%%%%%%%%%%%%%%%%%%%%%%%%%%%%%
\section{Results}
\label{Results}

\subsection{Acoustic spin transport}
\label{Acoustic_spin_transport}

The structure of the QWR samples is illustrated in Fig.~\ref{fig:Fig1}(a). As described in the Methods section, the GaAs QWRs are fabricated by combining steps of surface patterning and overgrowth by molecular beam epitaxy (MBE). In this process, the growth of an (Al,Ga)As/GaAs/(Al,Ga)As  QW stack over a patterned ridge leads to the formation of a thicker GaAs region (the QWR) at the ridge sidewalls, which is electrically connected to the QW\cite{PVS324}. 
The photoluminescence  features of the sample are summarized in  Fig.~\ref{fig:Fig1}(b). Here, the red and black curves compare PL spectra recorded under confocal excitation and detection on the QW region and on the ridge sidewall (corresponding to the QWR position), respectively. The former shows a single PL line with  a linewidth (full-width-at-half-maximum, FWHM) of \mev{4} associated with the electron-heavy hole QW exciton at \ev{1.546}. The spectrum recorded on the ridge sidewall shows the excitonic emission from the  QWR  at \ev{1.521}  with a FWHM of \mev{4.6} together with a second line at \ev{1.548} (FWHM of 5.4~meV). The latter stems from the QW regions near the QWR, which are slightly thinner than those further from the ridge sidewalls \cite{PVS324}.  The energy difference between these two lines yields a lateral confinement energy for electrons (holes) in the QWR of approximately \mev{22} (\mev{4}). 

%%%%%%%%%%%%%%%%%%%%%%%%%%%%%%%
% Figure 3: spin polarization data of QW
\begin{figure}[tbhp]
	\includegraphics[width=0.5\textwidth, angle=0, clip]{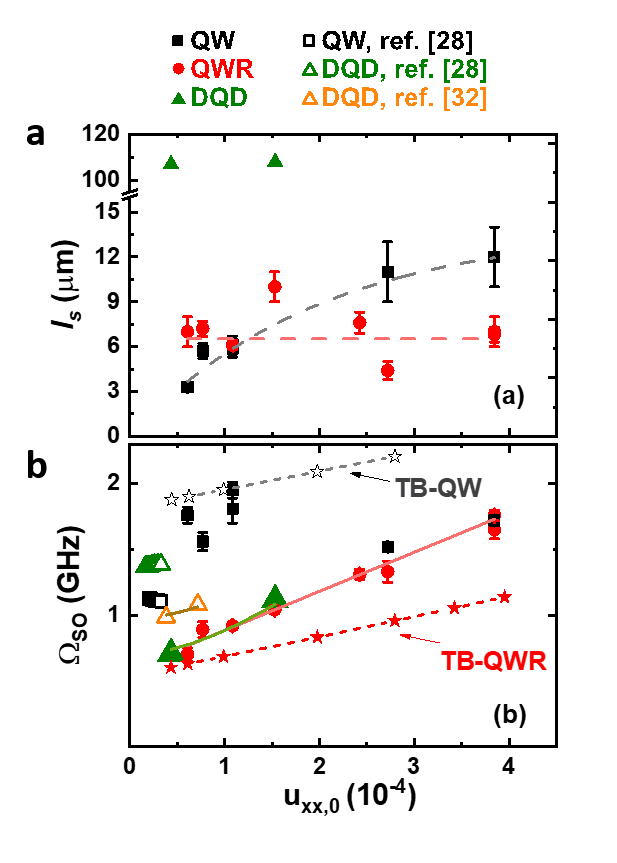}
	\vspace{ -0.5 cm}
	\caption{ 
		{\bf Spin dynamics under acoustic fields.}
		(a) Spin transport length, $\ell_s$,  and  (b) angular precession frequency, $\OSO$, as a function of the strain amplitude  $u_{xx,0}$  for transport along the  QW (black solid squares), QWR (red solid dots),  and DQDs on a 30~nm QW (green solid triangles). The orange triangles are for  DQDs in a 20~nm-thick QW~\cite{PVS199}.  The open symbols display data  for  transport by a single SAW  (open squares) and by DQDs (open triangles) in a 20~nm~thick QW\cite{PVS240}. Note that the $\OSO$ is scaled for the DQD data to account for the higher velocity. The open and solid ${\Large\star}$'s yield the $\OSO$ obtained from tight-binding (TB) calculations for the studied QWs and QWRs, respectively.	The dashed lines in (a) and (b) are a guide to the eye. The solid red and green lines in (b) are fits to  Eqs.~(\ref{EqSOA}) and (\ref{EqSOB}) of the QWR and DQD data, respectively.
	}
	\label{fig:Fig3}
\end{figure}
%%%%%%%%%%%%%%%%%%%%%%%%%%%%%%%

The optical detection of acoustically driven spin transport along the QWR is illustrated in Figs.~\ref{fig:Fig1}(c)-(e). The experiments were carried out using the geometry depicted in the upper right panel by exciting spins using a right-hand circularly polarized laser beam with energy below the QW resonance. The black and red profiles display the spatial distribution of the integrated PL from the QWR [detection window from \ev{1.517} to \ev{1.526}, cf.~Fig.~\ref{fig:Fig1}(a)] with right- and left-hand circular polarizations, respectively. In the absence of a SAW [Fig.~\ref{fig:Fig1}(c)], the emission  is restricted to the regions around the excitation spot and has a net right-hand circular polarization.  The PL profiles can be fitted with a Gaussian characterized by a FWHM of \micron{4}. The increased spatial PL spread compared to the size of the laser spot is attributed to the diffusion of the spin polarized carriers along the QWR axis. 

Figures~\ref{fig:Fig1}(d) and \ref{fig:Fig1}(e) display the corresponding profiles acquired in the presence of a SAW that captures the spin polarized carriers and transports them to trap centers at $x_\mathrm{QWR}=10~\mu$m, where the electrons and holes recombine. The trapping and recombination centers are defect regions along the QWR\cite{PVS324}.  The trapping mechanism, which leads to emission  at the same energy as the QWR, is attributed to carrier capture at centers at the interface between the QW (or QWR) layer and the (Al,Ga)As barrier layers assisted  by the transverse component of the SAW piezoelectric field, $F_z$~\cite{PVS324,PVS244}. These centers can capture carriers of one polarity during one half-cycle of the SAW and release them during the passage of carriers of the opposite polarity in the subsequent SAW half-cycle when $F_z$ reverses its sign. The centers provide efficient recombination centers to stop the transport and monitor the PL polarization. Note that the PL polarization depends on the transport distance $\Delta x$  changing from right- [Fig.~\ref{fig:Fig1}(e)] to left-hand circular polarization [Fig.~\ref{fig:Fig1}(d)] as $\Delta x$ increases from 5 to 10~$\mu$m.  This is due to the larger precession angle under the SO field accumulated while travelling a longer distance.

\subsection{Spin precession control}
\label{Spin precession control}

The procedure depicted in Figs.~\ref{fig:Fig1}(c)-(e) was applied to determine the spatial dependence of the spin polarization $\rho_s$ on the SAW amplitude. The panels of Fig. \ref{fig:Fig2} summarize $\rho_s$ profiles for acoustic transport in three geometries, which are illustrated in the corresponding upper panels: (a) along the QW, (b) along the QWR and (c) using DQDs.  For the QW and QWR, the transport distance $x$ is along the $[110]$-direction, but it is along the $[010]$-direction for the DQDs. In all cases, $\rho_s$ oscillates with a  period that reduces with increasing SAW amplitudes, which demonstrates the operation principle of the acoustic spin gate.  In particular, a distinct reversal of the spin polarization can be observed for both the QWR ($x = 12.5 \mu$m) and DQD ($x = 25 \mu$m) geometries, with the latter occurring well within the spin coherence length of transport. 

The solid lines are fits of the experimental data to an exponentially decaying cosine function of the form
\begin{equation}
\label{eq:spinpolQWR}
\rho_{s}(x)=\rho_{s}(0) \cos \left(\frac{\Omega_{SO}}{v_{SAW}} \Delta x\right) e^{- \Delta x/\ell_s}.
\end{equation}

\noindent The fit parameters are the total SO angular precession frequency $\Omega_\mathrm{SO}$  as well as the characteristic spin transport length $\ell_{s}$. $\Omega_\mathrm{SO}$ can be expressed in terms of the  oscillation period $L_{SO}=2\pi \vSAW/\Omega_{SO}$. The fit parameters $\ell_s$ and  $\Omega_\mathrm{SO}$ are displayed as filled symbols in Fig.~\ref{fig:Fig3} as a function of the SAW amplitude (stated in terms of the amplitude of the uniaxial strain component $u_{xx,0} \propto \sqrt{\PSAW}$, see definition below). For the QWR (solid red dots), $\ell_{s} \sim 7~\mu$m is independent of the SAW amplitude, and this value is equal to the one measured in the absence of a SAW\footnotePVS{see Supplemental Material, Sec.~SM2 at URL for details about the determination of the spin transport length\label{FNSM2}}. In contrast, $\ell_{s}$ for the QW (solid black squares) increases with SAW power [cf.~Fig.~\ref{fig:Fig3}(a)]--a behaviour which will be further addressed below. For the DQDs (solid green triangles), $\ell$ is considerably larger and comparable to the maximum measured transport distance. 

A further remarkable result is the strong dependence of \(\OSO\)  on the SAW amplitude [cf.~Fig.~\ref{fig:Fig3}(b)] for transport both along the QWRs and by DQDs. These structures thus act as acoustically controlled spin logic and polarization modulators, where the gate is determined by the acoustic amplitude. Their behaviour contrasts with that of the QW data as well as with previous acoustic transport studies of QWs and DQD on thinner QWs (open symbols in Fig. \ref{fig:Fig3}~\cite{PVS158,PVS240}), where the SAW-induced changes in \(\Omega_\mathrm{SO}\) are typically less than 10\%.

\subsection{SAW-related spin-orbit fields}
\label{sofields}

\begin{figure}[t!bhp]
	\includegraphics[width=0.90\columnwidth, angle=0, clip]{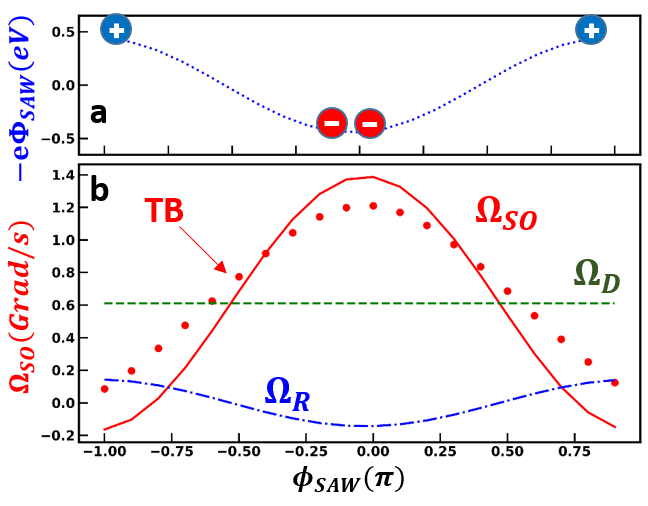}
	\caption{ {\bf Tight-binding calculations of the spin precession rate.} 
	SAW phase ($\pSAW$) dependence of the (a) electronic piezoelectric energy, $-e\FSAW$, and of the (b) electron-spin precession frequency determined by the tight-binding approach for a QW with the same thickness as the QWR under $\PSAW=100$~W/m. The dashed ($\Omega_\mathrm{D}$) and solid lines ($\Omega_\mathrm{R}$) are the corresponding Dresselhaus and piezoelectrically induced Rashba contributions, respectively, obtained from Eq.~(\ref{EqSOA}). 
}
\label{fig:Fig4}
\end{figure}

In order to quantify the SO fields induced by the SAW, we first note that the strain field of a Rayleigh SAW along \({\bf\vec x}||[110]\) consists of two uniaxial strain components \(u_{xx}=u_{xx,0}\cos{(\pSAW)}\) and  \(u_{zz}=u_{zz,0}\cos{(\pSAW)}\) as well as a phase-shifted shear component \(u_{xz}=u_{xz,0}\sin{(\pSAW)}\)\cite{PVS156}. Here, ${\bf \vec z}||[001]$ is the QW growth axis, $\pSAW=(\kSAW x -\wSAW t)$ is the SAW phase, and $ \kSAW= 2\pi/\lSAW$ and  $\wSAW$ are the SAW wave vector and angular  frequency, respectively. During  transport, the SAW piezoelectric potential $\FSAW=\Phi_{SAW,0}\cos(\pSAW)$ captures electrons around the SAW phase $\pSAW=0$ [cf. Fig.~\ref{fig:Fig4}(a)] with a strain field ($u_{xx}+u_{zz}$), where they are also subjected to a transverse piezoelectric field  $F_z$ proportional to $u_{xx}$\footnotePVS{see Supplemental Material, Sec.~SM5 at URL for details about the contributions of the SAW strain fields to the spin-orbit interaction\label{FNSM5}}.

%%%%%%%%%%%%%%%%%%%%%%%%%%%%%%%
% Figure 4: simulations
%%%%%%%%%%%%%%%%%%%%%%%%%%%%%%%

The lateral dimensions of the QWRs and DQDs are much larger  than the  QWR (or QW) thickness. For the determination of the $\OSO$ in QWRs and DQDs, we will neglect the role of the lateral confinement (and of the lateral interfaces) and assume that $\OSO$ is essentially equal to the one in a QW of the same thickness. 
We now address the different SO mechanisms acting on spins acoustically transported along the $x$-direction of a QW with the SAW velocity $\vSAW$. The effective electron wave vector  $k_x = m^*\vSAW / \hbar$ is determined by the effective electron mass $m^*$  ($\hbar$ is the reduced Planck's constant)\footnotePVS{see Supplemental Material, Sec.~SM3 at URL for details about the contributions of the SAW strain fields to the spin-orbit interaction\label{FNSM3}}. The bulk inversion asymmetry (BIA) of the III-V lattice  induces an intrinsic SO field (the Dresselhaus term\cite{Dresselhaus55a}), which leads to the precession of the moving spins with an angular frequency $\hbar{\Omega_\mathrm{D}} \sim \gamma k_x \left(\pi/w_z\right)^2$. Here,  $w_z$ denotes the extension of the electron wave function along $z$.
The structural inversion asymmetry (SIA) induced by the SAW gives rise to precession components  related to the strain field, $\hbar{\Omega_\mathrm{S}} =  \frac{1}{2} C_3 u_{xx}  k_x$ \cite{PVS272}, as well as to piezoelectric field $F_z$,  $\hbar{\Omega_\mathrm{R}}=  2r_{41} F_z k_x$. 
The terms $\gamma$, $r_{41}$, and $C_3$ are material parameters quantifying the strength of the different SO contributions (cf. Sec.~SM3).
The previous expressions can be combined such that the spin precession frequency for transport by a single SAW beam ($\Omega^{(QW)}_\mathrm{SO}$) and by DQDs ($\Omega^{(DQD)}_\mathrm{SO}$) can be stated as: \cite{PVS272,PVS240}

\begin{eqnarray}
	\Omega^{(QW)}_\mathrm{SO} &=&  \Omega_\mathrm{D} {\bf \hat y} - ( \Omega_\mathrm{R} + \Omega_\mathrm{S} ){\bf \hat y} 	\label{EqSOA} \\
	\Omega^{(DQD)}_\mathrm{SO}& =& \sqrt{2} \Omega_\mathrm{D} {\bf \hat x'}  - 2( \Omega_\mathrm{R} + \Omega_\mathrm{S} ) {\bf \hat y'} \label{EqSOB}
\end{eqnarray}

\noindent 
 In Eq.~(\ref{EqSOB}), ${\bf \hat x' = x-y}$ and ${\bf \hat y'=x+y}$ are the unit vectors parallel and perpendicular to the DQD propagation while \(\Omega_\mathrm{R}\) and  \( \Omega_\mathrm{S}\) refer to precession frequencies for a single SAW beam along ${\bf \hat x}$. The $\sqrt{2}$ term in this equation arises from the fact that the  propagation velocity of the DQDs equals to the vector sum of the velocities of the individual SAW beams.

Since $F_z\propto u_{xx}$ (hence $\Omega_\mathrm{R} \propto \Omega_\mathrm{S}$), Eq.~(\ref{EqSOA}) predicts a linear dependence of  $\Omega^{(QW)}_\mathrm{SO}$ on the SAW amplitude. For a quantitative comparison, we first note that while the $\gamma=$ \SI{17\pm2e-30}{\electronvolt \metre \cubed} \cite{PVS158} and $r_{41}$ = -\SI{59\pm7e-21}{\elementarycharge\metre\squared} \cite{PVS272}  have well-established values, the reported values of the strain-related parameter $C_3$ span a wide range (e.g., from  \SI{0.81}{\electronvolt\nano\metre} in n-type doped bulk GaAs \cite{Pikus88a,Beck05a} to \SI{0.31}{\electronvolt\nano\metre} in a GaAs(001) QW \cite{English_PRB84_155323_11}). 

The red solid line in Fig.~\ref{fig:Fig3}(b) is a fit of the QWR data to Eq.~(\ref{EqSOA}) and is then used to determine  $C_3=$\SI{-2.6\pm0.1}{\electronvolt\nano\metre} (details in Sec.~SM5). While  larger than previously reported values, this $C_3$ value reproduces very well the measured data for DQDs (after considering the scaling of $\OSO$ due to the larger $v_\textrm{DQD}$) and shows more reasonable agreement with the dashed lines in Fig.~\ref{fig:Fig3}(b) from tight-binding calculations. The relative signs of the terms of Eq.~(\ref{EqSOA}) is also consistent with  the distinctive slope of the red solid line in Fig.~\ref{fig:Fig3}(b).  These terms assumed that the electron spins were experiencing the maximum $\OSO$ as calculated in Fig.~\ref{fig:Fig4}(b), which supports the assumption that the electrons have a phase purity near $\pSAW=0$.

\subsection{Tight-binding calculations of the spin splittings}

In order to verify the predictions of the analytical model, we have also calculated the SO field $\OSO$ in QWs using the tight-binding (TB) method~\cite{PVS046,PVS060}.  The dots in Fig.~\ref{fig:Fig4}(b) display the $\OSO$ dependence on $\pSAW$ calculated for the QWR under $\PSAW=100$~W/m.  
In the TB calculations, the QWRs were modelled as a thicker QW, thus neglecting the effects of the weak lateral confinement.
The total SO component $\OSO$ oscillates around an average value $\Omega_D$ following the harmonic dependence of the SAW fields. To compare with the experiments, we further assume that the electrons are stored around the phases $\pSAW=0$ of minimum electronic piezoelectric energy $-e\FSAW$, as illustrated in Fig.~\ref{fig:Fig4}(a).  The SAW amplitude dependence of $\OSO$ at these phases calculated by the TB method  is displayed by the $\star$'s in Fig.~\ref{fig:Fig3}(b)  for the QWR and QW widths. The calculated linear dependence of $\OSO$ on strain amplitude, which corresponds to the parameter $C_3$ in Eq.(\ref{EqSOA}), does not depend on the thickness of the structures and is, thus, similar for the QW and QWR. The sign of the $C_3$ parameter determined by the TB method agrees with the experiments, thus reproducing the increase of the precession rate with SAW amplitude. Its magnitude $|C_3|=1.65$~eVnm  is slightly (33\%) smaller than the experimentally determined one, but still  well above previously  published experimental values.

%%%%%%%%%%%%%%%%%%%%%%%%%%%%%%%
% Figure 5: simulations
\begin{figure}[t!bhp]
	\includegraphics[width=1\columnwidth, angle=0, clip]{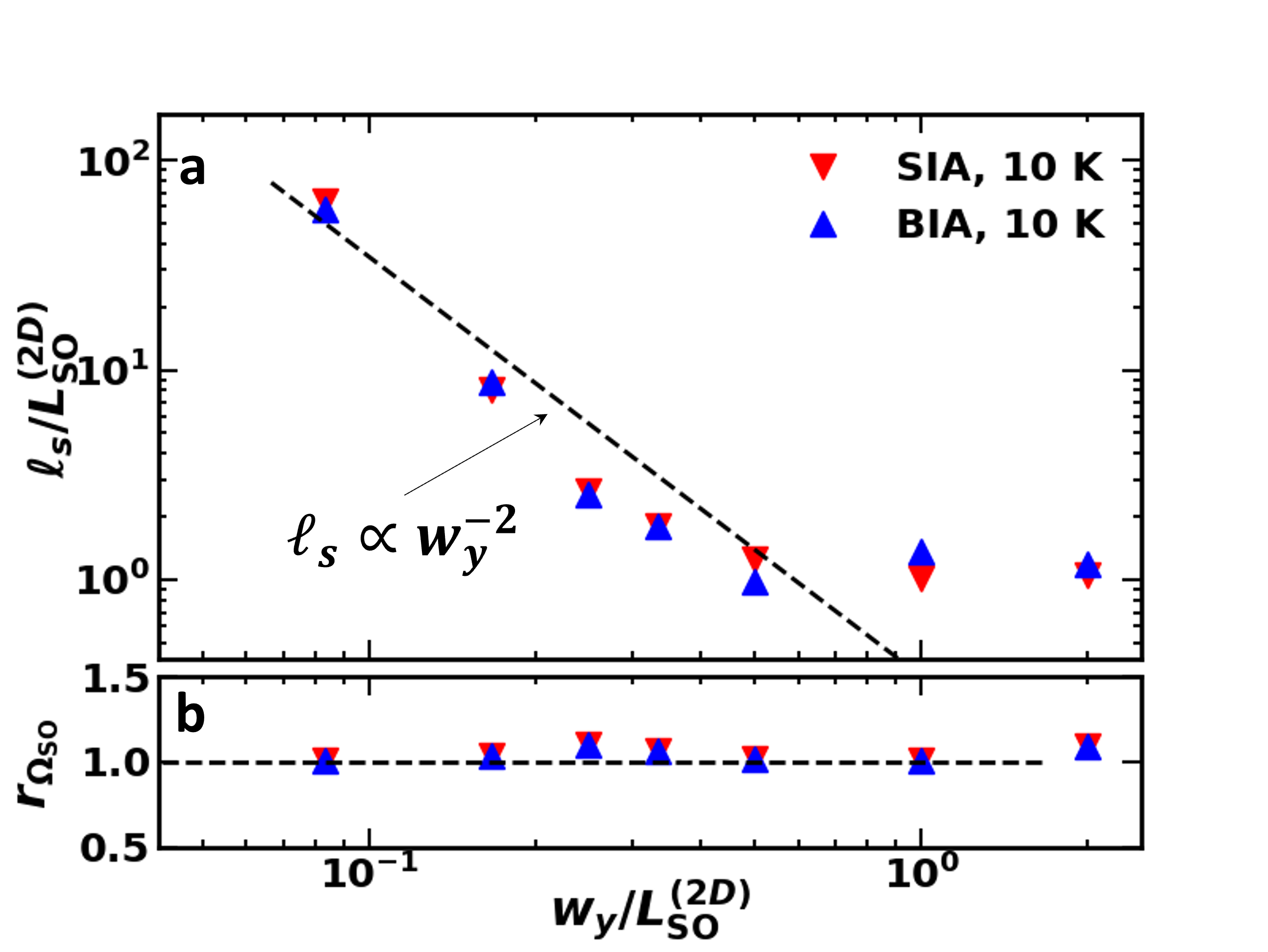}
	\caption{  
	{\bf Dimensionality effects on spin dephasing.} 
	(a)  Spin transport length, $\ell_s$,  and   (b) precession frequency ratio 	$r_{\OSO}=\OSO(w_y) /\OSO(w_y \rightarrow \infty)$ as a function of the  channel width, $w_y$, as determined from Monte-Carlo simulations of the acoustic spin transport under SO fields with BIA and SIA symmetries.
	The simulations were carried out assuming a mobility $\mu=4 m^2/(V s)$ and  temperature $T=10$~K.  
}
\label{fig:Fig5}
\end{figure}
%%%%%%%%%%%%%%%%%%%%%%%%%%%%%%%

\subsection{Dimensionality effects on the spin dynamics}
\label{dimension}

One interesting observation in connection with Figs.~\ref{fig:Fig3} and \ref{fig:Fig4}(b) is that the spin transport lengths in the QWRs are much shorter  than the ones in  the DQDs, despite the much smaller lateral confinement dimensions in the former structures. To understand the impact of the lateral confinement and motional narrowing effects on the spin dynamics, we have carried out Monte-Carlo simulations of the acoustic spin transport for channels with different widths $w_y$. The results are summarized in Fig.~\ref{fig:Fig5}. As expected from motional narrowing in the regime of 1D spin transport  ($w_y<\LSO^\mathrm{(2D)}$), $\ell_s$ increases with decreasing channel width as $ \ell_s \propto w^{-n_w} $ with $ n_w = 2 $~\cite{Kiselev_PRB61_13115_00}. In addition, we observe similar dependencies of $ \ell_s$ on the channel width for BIA and SIA SO fields. 
 
By using \(\LSO^\mathrm{(2D)}=\ell_s\sim 12~\mu\)m (as for the QW in Fig.~\ref{fig:Fig3}), the enclosure of electrons in  DQDs defined by  \(\lSAW=5.6~\mu\)m should increase \(\ell_s\) by a factor of \( (2\LSO^\mathrm{(2D)}/\lSAW )^2 \sim 18\). Such a mechanism qualitatively explains the long spin transport lengths by DQDs. It also
predicts  huge spin lifetimes and  transport lengths for the QWRs, which are, however, not observed in the experiments.  A possible explanation lies on the nature of the lateral confinement potential: while purely electrostatic for DQDs, it is imposed by structural interfaces in the QWRs. Spin dephasing due to frequent scattering events at these interfaces (known as  Elliot-Yafet processes\cite{Elliott54a}), as previously postulated for QWRs,\cite{Holleitner_PRL97_036805_06,PVS192} as well as interface-related SO fields may thus limit $\ell_s$ in structural channels.

A further remarkable experimental result is the weak SAW amplitude dependence of the spin precession rate $\OSO$ for transport in QW structures as compared to QWRs and DQDs [cf. Figs.~\ref{fig:Fig3}(a) and \ref{fig:Fig4}(a)]. This  behaviour in QWs also contrasts with the TB predictions as well as with the model leading to Eq.~(\ref{EqSOA}). 
Furthermore, it is at odds with the Monte-Carlo simulations  illustrated in Fig.~\ref{fig:Fig5}(b), which yield the dependence of the spin precession length on channel width. Here, $r_{\OSO}=\OSO(w_y) /\OSO(w_y \rightarrow \infty)$ is the ratio between the precession frequencies in a channel with finite width $w_y$ to the one in an unconstrained channel. One finds that $r_{\OSO}$ is essentially independent of the channel width. This result, which applies for SO fields of both BIA and SIA symmetries, can be qualitatively understood by taking into account that the spin precession rate $\OSO(w_y)$ along the SAW propagation direction ($x$) is determined by the carrier velocity component along $x$, which remains equal to  the SAW velocity as the carriers diffuse in the lateral direction~\footnotePVS{see Supplemental Material, Sec.~SM6 at URL for details about the dependence of the precession rate determined by the Monte-Carlo simulations on the channel width}.

We now briefly address a mechanism that can account for the weak dependence of $\OSO$ on SAW amplitude observed for the QW. This mechanism is based on fluctuations in the storage phase $\pSAW$ of the carriers in the SAW potential during the transport. 
The driving force for this transport is the longitudinal component of the SAW piezoelectric field given by $F_x=-e \partial \FSAW/\partial x$. In the previous sections, we have assumed that the electron spins remain at the SAW phase $\bar\phi_\mathrm{SAW}=0$ corresponding to the  minimum of the piezoelectric energy $-e\FSAW$, as illustrated in Fig.~\ref{fig:Fig4}(a). At these phases, however, $F_x$ vanishes. To sustain a steady-state motion at the SAW velocity,  both electrons (superscript $e$) and holes ($h$) must concentrate around a phase $\bar\phi^{(i)}_\mathrm{SAW}$ ($i=e, h$) of the SAW potential satisfying $\sin(\bar\phi^{(i)}_\mathrm{SAW}) = r^{(i)}_d= \vSAW/(\mu^{(i)} F_z)$ [rather than at $\pSAW=0$, cf. Fig.~\ref{fig:Fig4}(a)]. 
At these phases, the SAW-induced SO fields are reduced by a factor $\cos(\bar\phi^{(i)}_\mathrm{SAW})=(1-(r^{(i)}_d)^2)$ with respect to their maximum at $\pSAW=0$. As a result, the effective spin precession frequency reduces and becomes dependent on the effective carrier mobility. Due to the narrower thickness $w_z$, the transport mobility $\mu^{(i)}$ in the QWs is expected to be lower than in the QWR and DQD cases  (the ambipolar mobility scales with  $w_z^{-6}$ in undoped QWs~\cite{Voros_PRL94_226401_05}), thus reducing the  SAW-induced SO fields relative to the Dresselhaus contribution.  As mentioned in connection with  Fig.~\ref{fig:Fig3}(a),  $\ell_{s}$ increases with SAW power for the spins in the QW, thus indirectly indicating a connection between the mobility and the spin dynamics. 
According to this mechanism, the lower efficiency of the spin gates in the QW arises from the lower ambipolar mobility relative to the QWR and the DQDs due to its smaller thickness.
Further studies are, however, required to quantify the impact of this effect.

\section{Discussion}
\label{discussion}

In this work, we have demonstrated the ability to acoustically transport and, simultaneously, control the polarization of electrons spins stored within moving potential dots defined by SAWs in GaAs QWs and QWRs. Within the flying control gates, the spins precess around a SAW-induced SO field, and the precession rate can be dynamically changed by a factor of over 2.5 by varying the SAW amplitude, which enabled the controlled flip of the spin polarization. The experimental results for the precession rates markedly exceed previous results and agree well with the predictions of an analytical model for the SAW-induced spin-orbit fields as well as with microscopic tight-binding calculations. The latter of which enable a precise determination of the strain related spin-orbit parameters. We have also addressed the mechanisms governing spin relaxation while considering the role of the carrier mobility and of the lateral interfaces on the spin dynamics. Here, the shorter spin transport lengths measured in structurally defined QWRs as compared to electrostatically defined DQDs are attributed to spin dephasing via frequent scattering at the lateral (Al,Ga)As interfaces. Details of the scattering mechanisms are, however, presently unknown and calls for additional structural studies of the QWR interfaces.

The high degree of dynamical spin control in QWRs and DQDs demonstrated here enables dynamic control by simply changing the acoustic amplitude of the carrier wave. An obvious and important advantage is that both spin transport and manipulation can be performed through this flying spin gate in a single structure (e.g., a QWR) without requiring  extra components (e.g., electrostatic gates for Rashba spin control). The experiments in Fig.~\ref{fig:Fig2}(b) directly illustrate an electro-optical polarization modulators based on this approach. Here, the circular polarization of the PL can be set to an arbitrary value by simply varying the SAW amplitude. This functionality can be naturally extended to the transport of single spins\cite{Bertrand_NN11_672_16} as well as for the generation of polarized single photons\cite{PVS152}, thus enabling the realization of a spin-based quantum information processor with a photonic interface. 

%%%%%%%%%%%%%%%%%%%%%%%%%%%%%%%%%%%%%%%%%%%%%%%%%%%%%%%%%%%%%%%%%%%%%%
% METHODS
%%%%%%%%%%%%%%%%%%%%%%%%%%%%%%%%%%%%%%%%%%%%%%%%%%%%%%%%%%%%%%%%%%%%%%

\section{Methods}
\label{Methods}

\subsection{QWR Sample Fabrication}
The planar QWRs used in this work were fabricated using MBE by overgrowing a 10~nm-thick QW  on a GaAs(001) substrate  pre-patterned with shallow (approximately 30~nm high) rectangular ridges [cf.~Fig.~\ref{fig:Fig1}(a)]~\cite{PVS324}. The anisotropic nature of the MBE growth induces a local thickening of the QW at the ridge sidewalls oriented along the ${\bf \vec{x}}||[110]$-direction along the surface, thus forming a QWR parallel to this sidewall. The thickness and width of the QWR are \nanometer{25\pm5} and \nanometer{200\pm5}, respectively, as determined by scanning transmission electron microscopy~\cite{PVS324}. The moving potential dots for carrier and spin transport are created by propagating a SAW  with a wavelength $\lSAW=4~\mu$m (frequency of $726$~MHz at \SI{15}{\kelvin} and velocity $\vSAW=2904$~m/s)  along the wire axis. The SAW is generated by a split-finger interdigital transducer deposited on the sample surface. Its amplitude will be quantified in terms of the linear acoustic power density $\PSAW$, defined as the  ratio between the coupled acoustic power and the width of the SAW beam. During the acoustic transport, the carriers are confined within dots with dimensions equal to the QWR width  and less than $\sim\lSAW/2$ along the directions perpendicular and parallel to the SAW propagation, respectively.  Acoustic spin transport was also investigated in  the QW embedding the QWR. In this case, carrier motion is unconstrained in the direction perpendicular to the transport. 

\subsection{DQD Sample Fabrication}
The DQDs were created via the interference of two orthogonal SAW beams propagating along the ${\bf \vec{x}}||[110]$ and ${\bf \vec{y}}||[\bar 1 10]$ directions of a 30~nm thick QW on GaAs(001)~\cite{PVS152}. The interference of the piezoelectric fields of the SAWs creates an array of DQDs propagating along the ${\bf \vec{x^\prime}}||[010]$ surface direction~\cite{PVS152}. DQDs yield the longest, so far, reported acoustic spin transport distances~\cite{PVS152,PVS165}.  SAWs with a wavelength of  $\lSAW=5.6~\mu$m were used and yield DQDs with dimensions of approximately $\lSAW/2 \times \lSAW/2$ propagating with a velocity $v_\textrm{DQD} = \sqrt{2} \vSAW = $ 4115~m/s.

\subsection{Photoluminescence measurements}
The spectroscopic photoluminescence (PL) studies of the spin transport were performed at low temperatures (10-20~K) in a microscopic PL setup with radio-frequency (rf) wiring for SAW excitation. The spins were optically excited using a circularly polarized laser beam with tunable wavelength ($\lambda_L$ between 760 and 808~nm) focused onto a $\sim 2~\mu$m wide spot on the sample surface. The PL emitted along the SAW path with left- ($I_L$) and right-hand  ($I_R$) circular polarization was collected with spatial resolution and used to determine the spin polarization~\cite{Dyakonov_Springer-Verlag_08} $\rho_s=({I_{R}-I_{L}})/({I_{R}+I_{L}})$~\footnotePVS{see Supplemental Material, Sec.~SM1 at URL for further details about the experimental determination of the spin polarization\label{FNSM1}}. 
Since hole-spin relaxation in  QWs and QWRs is typically much faster than the one for electrons \cite{Baylac_SS_326_161_95},  $\rho_s$ essentially reflects the electron spin dynamics.

\subsection{Tight-binding Calculations}
% Tight-binding: moved
The analytical model for the SO fields applies for a QW with infinite potential barriers and, thus, neglects the effects of the QW interfaces. In addition, the model neglects the impact of the SAW field on the electronic states, which can be significant for the valence band states in wide QWs~\cite{PVS139}. The the SO field  $\OSO$ contributions were calculated in QWs using the tight-binding (TB) method~\cite{PVS046,PVS060}. The effects of the SAW were taken into account by using the strain and piezoelectric fields to determine the atomic positions and on-site potentials, respectively, within the TB supercell~\footnotePVS{see Supplemental Material, Sec.~SM4 at URL for further details about the tight-binding calculation procedure\label{FNSM4}}. %Note that the calculations are without free parameters. 

\subsection{Monte Carlo spin dynamics}
In the calculations, the spins are assumed to move along the $x$-direction with the SAW velocity while having a random motion along the $y$-direction with mean-free-path $\ell_p$ and thermal velocity $v_p$ defined by the electron mobility $\mu$  and  temperature~\footnotePVS{see Supplemental Material, Sec.~SM6 at URL for details about the Monte-Carlo simulations\label{FNSM6}}. Figure~\ref{fig:Fig5}(a) and (b) display the simulated dependence of $\ell_s$ and $\OSO$ for transport in channels with different widths $w_y$ under BIA and SIA SO fields. The SO fields were assumed to have an amplitude $\OSO^\mathrm{(2D)}=2\pi\vSAW/\LSO^\mathrm{(2D)}$ dictated by a  spin precession period $L_\mathrm{SO}^{(2D)}=12~\mu$m similar to the ones measured for QWs in Fig.~\ref{fig:Fig2}. We further assumed   $\mu = 4 m^2/(V s)$ and  $T=4$~K, which yield $\ell_p=0.1~\mu$m.

\section{Author Contributions}
\label{Author_Contributions}

P.L.J.H. performed the experiments and analysis on the QW and QWR samples, and J.A.H.S. performed the experiments and analysis using the DQDs.  K.B. performed the MBE growth of the samples. Initial discussions and experimental design and development between H.S., Y.K., and P.L.J.H. were critical inputs to the presented results.  P.V.S. led the theory calculations and supervised the project.  P.L.J.H., P.V.S., and J.A.H.S. wrote the manuscript with input from the other authors.

%%%%%%%%%%%%%%%%%%%%%%%

\begin{acknowledgments}
We thank S. Ludwig for discussions and for a critical reading of the manuscript, {A.C.} Poveda for the help with the graphics, as well as S. Rauwerdink  for the help in the fabrication of the samples. J. Stotz would also like to acknowledge the Natural Science and Engineering Research Council of Canada and the Alexander von Humboldt Foundation.  This project has received funding from the European Union's Horizon 2020 research and innovation program under grant agreement No 642688.
\end{acknowledgments}

\end{document}

\title{Supplementary Material:\\
Flying electron spin control gates}% Force line breaks with \\

\author{Paul L. J. Helgers} 
\affiliation{Paul-Drude-Institut f{\"u}r Festk{\"o}rperelektronik, Leibniz-Institut im Forschungsverbund Berlin e.V., Hausvogteiplatz 5--7, 10117 Berlin, Germany }
\affiliation{NTT Basic Research Laboratories, NTT Corporation, 3--1 Morinosato-Wakamiya, Atsugi, Kanagawa 243-0198, Japan}

\author{James A. H. Stotz} 
\affiliation{Paul-Drude-Institut f{\"u}r Festk{\"o}rperelektronik, Leibniz-Institut im Forschungsverbund Berlin e. V., Hausvogteiplatz 5--7, 10117 Berlin, Germany }
\affiliation{Department of Physics, Engineering Physics \& Astronomy, Queen’s University, Kingston, ON, K7L 3N6 Canada}

\author{Haruki Sanada}
\author{Yoji Kunihashi}
\affiliation{NTT Basic Research Laboratories, NTT Corporation, 3--1 Morinosato-Wakamiya, Atsugi, Kanagawa 243-0198, Japan}

\author{Klaus Biermann}
\author{Paulo V. Santos}
\email{santos@pdi-berlin.de} 
\affiliation{Paul-Drude-Institut f{\"u}r Festk{\"o}rperelektronik, Leibniz-Institut im Forschungsverbund Berlin e. V., Hausvogteiplatz 5--7, 10117 Berlin, Germany }

\date{\today}% It is always \today, today,

%\keywords{polaritons; OPO; SAW}%Use showkeys class option if keyword display desired
\maketitle

%\tableofcontents

%%%%%%%%%%%%%%%%%%%%%%%%%%%%% Supplementary Information %%%%%%%%%%%%%%%%%%%%%%%%%%%%%%%%%%%%
%\clearpage
%\appendix
%\widetext 

\beginsupplement

%%%%%%%%%%%%%%%%%%%%%%%%%%%%%%%%%
% SISpinPol
%%%%%%%%%%%%%%%%%%%%%%%%%%%%%%%%%
\section{Spin polarization measurements}
\label{SISpinPol}

The spectroscopic photoluminescence (PL) studies of the spin transport were carried out in a helium flow cryostat (10-20~K) with optical access and radio-frequency wiring for SAW excitation. The spins were optically excited using a circularly polarized laser beam from a tunable Ti-sapphire laser (wavelengths $\lambda_L$ between 760 and 808~nm) focused onto a $\sim~2~\mu$m wide spot on the sample surface by a 20x objective [L in Fig.~2(c) of the main text].  The same objective collects the PL emitted from along the SAW path and directs it to a triple-grating spectrometer operating in the subtractive mode and equipped with a cooled CCD detector. Spatially resolved PL maps are recorded by imaging the acoustic transport path onto the input slit of the spectrometer. An optical arrangement consisting of a quarter wave plate and a beam polarization displacer  placed in front of the spectrometer slit  shifts the PL images  with right ($I_R$) and left ($I_L$)  hand circular polarizations along the slit direction. In this way, we simultaneously record PL images of the spatial distribution of the two polarizations on the CCD. 

The spin polarization $\rho_s$ during acoustic transport can be determined by exciting spins with a right hand circularly polarized laser beam and detecting the PL intensities $I_R$ and $I_L$ according to:

\begin{equation}
\rho_s=\frac{I_R-I_L}{I_R+I_L}.
\end{equation}

\noindent Since hole-spin relaxation in  QWs and QWRs is typically much faster than for electrons ~\cite{Baylac_SS_326_161_95}, we assume that $\rho_s$ reflects only the electron spin dynamics.

In order to correct for artifacts arising from dichroism of the optical components of the setup, the polarization experiments were  carried out by recording PL maps excited with   a right-hand and then  a left-hand circularly polarized laser beam. We will denote the corresponding PL intensities as  $I_{LL}$, $I_{LR}$, $I_{RL}$ and $I_{RR}$, where the first (second) index corresponds to the circular polarization of the excitation (PL emission).  For an ideal optical system (i.e., with transmission independent of the polarization), one expects a ratio

\begin{equation}
a_r^2=\frac{I_{LL}I_{RL}}{I_{RR}I_{LR}}=\frac{t_L}{t_R}
\end{equation}

\noindent equal to unity. The presence of dichroic elements in the optical path makes $a_r\neq1$. The effect of these components on the spin polarization can be corrected by using the following expression to determine the spin polarization:

\begin{equation}
\rho_s=\frac{a_rI_{RR}-I_{RL}}{a_rI_{RR}+I_{RL}}.
\end{equation}

%%%%%%%%%%%%%%%%%%%%%%%%%%%%%%%%%
% SILifetimes
%%%%%%%%%%%%%%%%%%%%%%%%%%%%%%%%%
\section{Spin lifetimes}
\label{SILifetimes}

% --------------------------------------------------------------------------------
\begin{figure}
	\includegraphics[width=0.75\columnwidth]{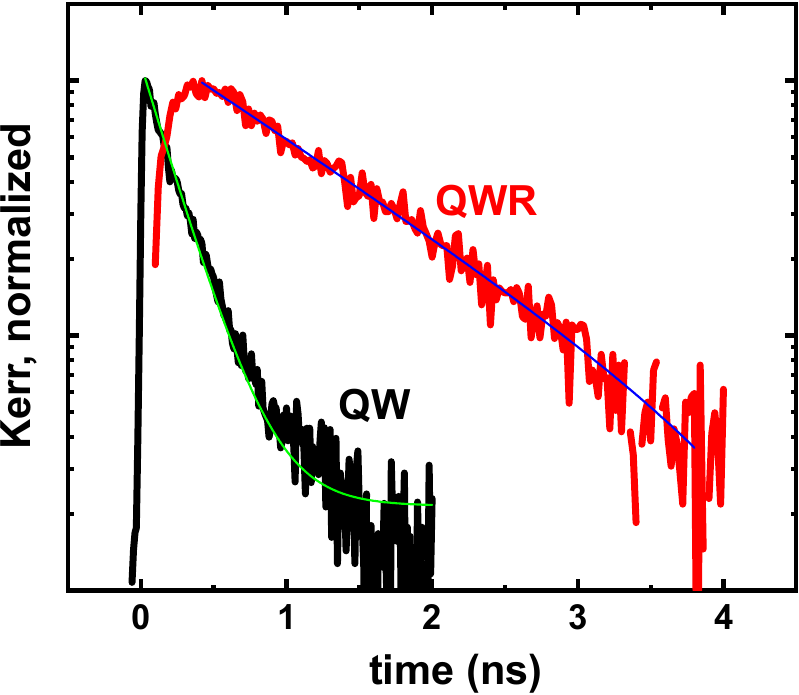}
	\caption{Time-resolved Kerr polarization rotation measured on the QW (black) and QWR (red). The lines are exponentially decaying functions yielding a Kerr rotation decay time of approximately \SI{200}{\pico\second} and \SI{1}{\nano\second} for the QW and the QWR, respectively.}
	\label{fig:SI-Kerr}
\end{figure}
% --------------------------------------------------------------------------------
 
The spin lifetimes $\tau_s$ in the absence of acoustic excitation were determined by time-resolved Kerr rotation measurements carried out in the  pump-probe configuration~\cite{PVS240}. Typical temporal Kerr traces for   the QW (black) and  QWR (red) are shown in Fig.~\ref{fig:SI-Kerr}. The measurements were performed using an excitation energy of \SI{1.551}{\electronvolt} and power density of \SI{8}{\watt\per\centi\metre\squared}. The traces can be fitted with an exponential decaying function yielding Kerr rotation decay   times $\tau_K=\SI{200}{\pico\second}$ for the QW and $\tau_K= \SI{1}{\nano\second}$ for the QWRs. 

The spin decay time was determined by combining the Kerr rotation decay times with the carrier recombination lifetimes obtained from time resolved PL measurements. This procedure yields  spin lifetimes of $\tau_s=\SI{0.5}{\nano\second}$ and $\tau_s=\SI{2}{\nano\second}$ for the QW and QWRs, respectively. The much longer spin lifetime in the QWR as compared to the QW is attributed to (i) the lateral confinement of the QWR carriers by the surrounding QW, leading to motional narrowing and (ii) the larger thickness of the QWR. Whereas both effects suppress D'yakonov-Perel' spin relaxation~\cite{Kiselev_PRB61_13115_00,PVS272}, the latter also  decreases of the electron-hole exchange interaction and its associated spin relaxation ratio \cite{MSS93a}. 

%%%%%%%%%%%%%%%%%%%%%%%%%%%%%%%%%
% SAW-related SO fields
%%%%%%%%%%%%%%%%%%%%%%%%%%%%%%%%%

\section{SAW-induced SO fields}
\label{SISAWSO}

In this section, we briefly summarize the SO mechanisms related to the strain and piezoelectric fields induced by a SAW. In the absence of a SAW, the conduction band spin splitting for electrons moving along the $\langle 110 \rangle$-directions of an intrinsic GaAs QW on a (001) substrate is given by the Dresselhaus term:

\begin{equation}
\Omega_\mathrm{D} \sim \pm \gamma k \left[ \left(\frac{\pi}{w_z}\right)^2   - \frac{1}{2}k^2   \right],
\label{dEeqD}
\end{equation}
\noindent %where The  Dresselhaus term is   given by Eq.~\ref{Eq100b} by assuming $\langle k^2_z=\pi/(d_{QW}+d_0)\rangle$, 
where $\gamma$ is a material constant and $w_z=d_{QW}+2 d_0$ is the total extension of the electronic wave function, assumed to be equal to the QW width ($d_{QW}$) plus  the penetration depth in the barrier layers ($d_0$). 
For an electron moving with the SAW velocity $\vSAW$, the wave vector $k$ is obtained from the electron momentum according to:

 % ------------------------------------------------
 \begin{equation}
 k= \frac{m^*\vSAW}{\hbar}.
 \label{EqSIk}
 \end{equation}
 % ------------------------------------------------
 
 \noindent where $m^*$ is the conduction band (CB) effective mass.

A uniaxial strain leads to a linear term in the CB splitting given by:\cite{Dyakonov83a,Pikus88a,Meier_84}
% (see also page 91 of Ref.~\onlinecite{Meier_84})

\begin{equation}
H_\mathrm{S}=\frac{1}{2}\left[C_3 (\hat\sigma \vec \phi)+C'_3 (\hat\sigma \vec
  \psi)\right]
\label{EqDyakonov1}
\end{equation}

\noindent where $\phi_z=\varepsilon'_{xz}k_z - \varepsilon'_{yz}k_y'$ and $\psi_z=k_z(\varepsilon'_{xx}-\varepsilon'_{yy})$ and cyclic permutations. The $\hat\sigma_i$ denote the Pauli matrices. The apostrophe (~$'$~) indicates that the strain components are relative to the conventional axes $x'$, $y'$, and $z'$.  This Hamiltonian gives rise to a strain-induced spin  precession frequencies given by~\cite{Beck05a}

% ------------------------------------------------
\begin{equation}
\hbar\Omega_\mathrm{S} = C_3
\left[ 
\begin{array}{c}
\varepsilon'_{yx} k_y - \varepsilon'_{zx} k_z\\
\varepsilon'_{zy} k_z - \varepsilon'_{yx} k_x\\
\varepsilon'_{xz} k_x - \varepsilon'_{yz} k_y\\
\end{array} \right] + C_3'\left[
 \begin{array}{c}
k_x (\varepsilon'_{yy}- \varepsilon'_{zz})\\
k_y (\varepsilon'_{zz}- \varepsilon'_{xx})\\
k_z (\varepsilon'_{xx}- \varepsilon'_{yy})\\
\end{array} \right]
\label{EqDyakonov2}
\end{equation}
% ------------------------------------------------

In order to address the SAW strain field, it is convenient to use a reference frame with the $x$-axis along the SAW propagation direction, the $y$-axis on the sample surface and perpendicular to $x$, and the $z$-axis perpendicular to the surface. A SAW propagating along the $[110]$- or $[\bar1 1 0]$-directions of the (001) surface induces three non-vanishing engineering strain components $u_{xx}$, $u_{zz}$, and $u_{xz}$. Furthermore, we use $u_{ij}$ to denote the engineering strain components and, thus, to distinguish them from the physical strain ones 
 $\varepsilon_{i,j}$ ($\varepsilon_{ij}=u_{ij}$ for $i=k$ and $\varepsilon_{ij}=u_{ij}/2$ for $i \neq j$). 
 
In a piezoelectric material, the strain field from the SAW generates a piezoelectric polarization field D given by
\begin{equation}
D=e_{14}[ u_{xz}, 0, u_{xx}]^T,
\label{EqPolL1}
\end{equation}

\noindent where the superscript $T$ denotes the vector transposition operation. When the SAW  fields are transformed to the Cartesian reference frame, the  strain and electrical polarization fields become:

% ------------------------------------------------
\begin{eqnarray}
\varepsilon'&=&\left[u_{xx}/2,u_{xx}/2,u_{zz}, \pm u_{xz}/\sqrt{2},u_{xz}/\sqrt{2},\pm u_{xx}\right]^T,
\label{EqOSOSV} \\
D'&=& e_{14}[2 u_{xz}, 0, u_{xx}]^T,\label{EqPolL2}
\end{eqnarray}
% ------------------------------------------------

\noindent  where upper and lower signs apply for SAWs propagating along the [110]- \nPVS{($\theta=\pi/4$)} and $[\bar 1 1 0]$-directions\nPVS{ ($\theta=-\pi/4$)}, respectively. By  substituting Eq.~\ref{EqOSOSV} into Eq.~\ref{EqDyakonov2}
one obtains:

% ------------------------------------------------
\begin{equation}
  \hbar\Omega_\mathrm{S} = \frac{1}{2\sqrt{2}} k
  \left[C_3 u_{xx} + C_3' (u_{xx}-2u_{zz})\right]
  \left[ 
  \begin{array}{c}
  1\\
  -1\\
  0\\
  \end{array} \right] 
  \label{EqDyakonov3}
  \end{equation}
  % ------------------------------------------------
  
\noindent Note that the strain induces a SO component perpendicular to the SAW propagation direction.

A SAW in a piezoelectric material also induces  a  transverse piezoelectric field $F_z$ associated with $D$, which gives rise to a Rashba spin-orbit contribution. As a consequence, SAWs along a $\langle 110\rangle$  induce  three contributions for the CB spin splitting, all oriented in-plane and perpendicular to the SAW propagation direction, given by\nPVS{(mathematica/piezosaw/spin.nb)}:

% ------------------------------------------------
\begin{eqnarray}
\hbar \Omega_\mathrm{SO} &=& \Omega_\mathrm{D} + \Omega_\mathrm{S} + \Omega_\mathrm{R}  \label{dEeqT}\\
\Omega_\mathrm{S} &=& \mp  k \frac{1}{2}\left[C_3 u_{xx} + C_3'( u_{xx}-2u_{zz})\right]  \label{dEeqS}\\
\Omega_\mathrm{R} &=& -2 k r_{41} F_z  \label{dEeqR}
\end{eqnarray}
% ------------------------------------------------

\noindent 

The upper and lower signs apply for SAWs along the [110] \nPVS{($\theta=\pi/4$)} and $[\bar 1 1 0]$ directions\nPVS{ ($\theta=-\pi/4$)}, respectively. We will show in Sec.~\ref{SITBC3} that the $C_3'$ term in Eq.~\ref{dEeqS} is much smaller than the one associated with $C_3$. This term will be neglected in the main text and in the subsequent analyses.

\begin{table}
  \caption{ Tight-binding parameters used in the calculation of spin-orbit conduction band splitting.  The notation used corresponds to the one of Ref.~\protect\onlinecite{Lok93a}.   In order to account for the valence band discontinuity we subtracted 
  0.007~eV from the diagonal parameters (i.e., $E_s$, $E_p$, and 
  $E_\mathrm{s^*}$) of the Al$_\mathrm{0.15}$Ga$_\mathrm{0.85}$As barrier layers. 
   }
  \begin{tabular}{| l | c | c |}
  \hline
  Parameter 	&  GaAs\cite{Schulman85a} (eV)	& Al$_\mathrm{0.15}$Ga$_\mathrm{0.85}$As
  (eV)\\
  \hline
  $E_s$(anion) 	&	-8.457	&	-8.3172	\\
  $E_p$ (anion) 	&	0.9275	&	0.92938	\\
  $E_s$(cation) 	&	-2.7788	&	-2.53673	\\
  $E_p$ (cation) 	&	3.5547	&	3.55250	\\
  $V_\mathrm{ss}$ 	&	-6.4513	&	-6.48321	\\
  $V_\mathrm{xx}$ 	&	1.9546	&	1.95121	\\
  $V_\mathrm{xy}$ 	&	4.77	&	4.69110	\\
  $V_\mathrm{s_0p}$(anion,cation) 	&	4.48	&	4.57360	\\
  $V_\mathrm{s_1p}$(cation,anion) 	&	7.85	&	7.49390	\\
  $E_\mathrm{s^*}$(anion) 	&	8.4775	&	8.32833	\\
  $V_\mathrm{s_0^*p}$(anion,cation) &	4.8422	&	4.78967	\\
  $E_\mathrm{s_1^*}$(cation) 	&	6.6247	&	6.64005	\\
  $V_\mathrm{s^*p}$(cation,anion) &	7	&	6.69940	\\
  $3\lambda_a$ (anion) 	&	0.39	&	0.39171	\\
  $3\lambda_c$ (cation) 	&	0.174	&	0.15150	\\
  \hline
  \end{tabular}
  \label{TBTable}
  \end{table}
  % ********************************************************************

%%%%%%%%%%%%%%%%%%%%%%%%%%%%%%%%%
% Tight-Binding calculations
%%%%%%%%%%%%%%%%%%%%%%%%%%%%%%%%%
\section{Tight-Binding calculations of spin splittings in QWs}
\label{SITB}

We present in this section details about the calculations of the spin splitting of the conduction band of GaAs QWs under SAW fields. In bulk GaAs, this splitting results from a three-band coupling process involving the  s-like GaAs conduction band and the p-like conduction and valence bands~\cite{Cardona_PRB38_1806_88}. The three-band interaction process leads to the cubic dependence of the spin splitting on wave vector $k$. Calculation of the splitting requires, therefore, that one takes into account  interactions between of all these bands. As demonstrated in Ref.~\onlinecite{Cardona_PRB38_1806_88}, this can be carried out using ab-initio approaches or  empirical approaches using a large number of bands (e.g.,  a 14 band {$\bf {k\cdot p}$} method). Here, we determine the splittings using the empirical tight-binding (TB) method\cite{Slater54a} following the  approach described in Refs.~\onlinecite{PVS046,PVS060}. This atomistic approach  enables the determination of the band structure of QWs with widths up to a few tens of nanometres using moderate computational efforts. Although empirical in the sense that it uses parameters fitted to the bulk band structure, it can be regarded as ``microscopic'' when compared with other approaches such as the  {$\bf {k\cdot p}$} effective mass calculations. 

The TB calculations were carried out using a basis consisting of  $sp^3s^{*}$ orbitals~\cite{Vogl_JPCS44_365_83} including spin-orbit coupling\cite{Hass83a,Chadi77a}. This orbital basis consists of 10 orbitals per atom and includes only nearest-neighbour interactions.  It has been shown to reproduce very well the highest valence and the lowest conduction bands of most bulk semiconductors\cite{Vogl_JPCS44_365_83}. The tight-binding parameters employed  for the GaAs QW and the Al$_\mathrm{0.15}$Ga$_\mathrm{0.85}$As barrier layers are summarized in Table~\ref{TBTable}.%I of Ref.~\onlinecite{PVS330}. 

The calculations were performed for an electron with wave vector given by Eq.~\ref{EqSIk} in a periodic super-cell  consisting of a single QW  with sandwiched between 4.5~nm wide Al$_\mathrm{0.15}$Ga$_\mathrm{0.85}$As barriers.  The atomic positions within the unit cell were deformed in order to take into account the effects of the SAW strain field determined in the previous section. The effects of the  piezoelectric field $F_z$ were incorporated  in the calculations by adding the position-dependent electrostatic energy to the on-site TB parameters. 
 The following convention applies for the crystallographic orientation of the axes:

\begin{itemize}
  \item in the unit cell, the anion (As) sits at the origin while the cation (Ga) is at $(a_0/4)(111)$ (this is the same as in Ref.~\onlinecite{Cardona_PRB38_1806_88});
  \item a positive (piezo) electric field corresponds to a field along the growth direction.
\end{itemize}

\subsection{Tight-binding calculation of the spin-orbit strain coefficients}
\label{SITBC3}

As discussed in Sec.~\ref{SISAWSO}, the SAW strain induces SO contributions described by two material constants $C_3$ and $C_3'$. 
Reported values for these contributions span a wide range:  
$C_3=0.47$~eVnm~\cite{Pikus84a}, $C_3=0.52$~eVnm~\cite{Dyakonov83a}, and $C_3=0.81$~eVnm~\cite{Beck05a}. $C'_3$ has not been measured (nor  calculated), and should be negligible in comparison with $C_3$.%~\cite{Beck_thesis}.%,Pikus84a} 
In this section, we use the TB approach to estimate the  magnitude of these contributions. 

\begin{figure}[tb]

\includegraphics[width=1.\columnwidth,clip]{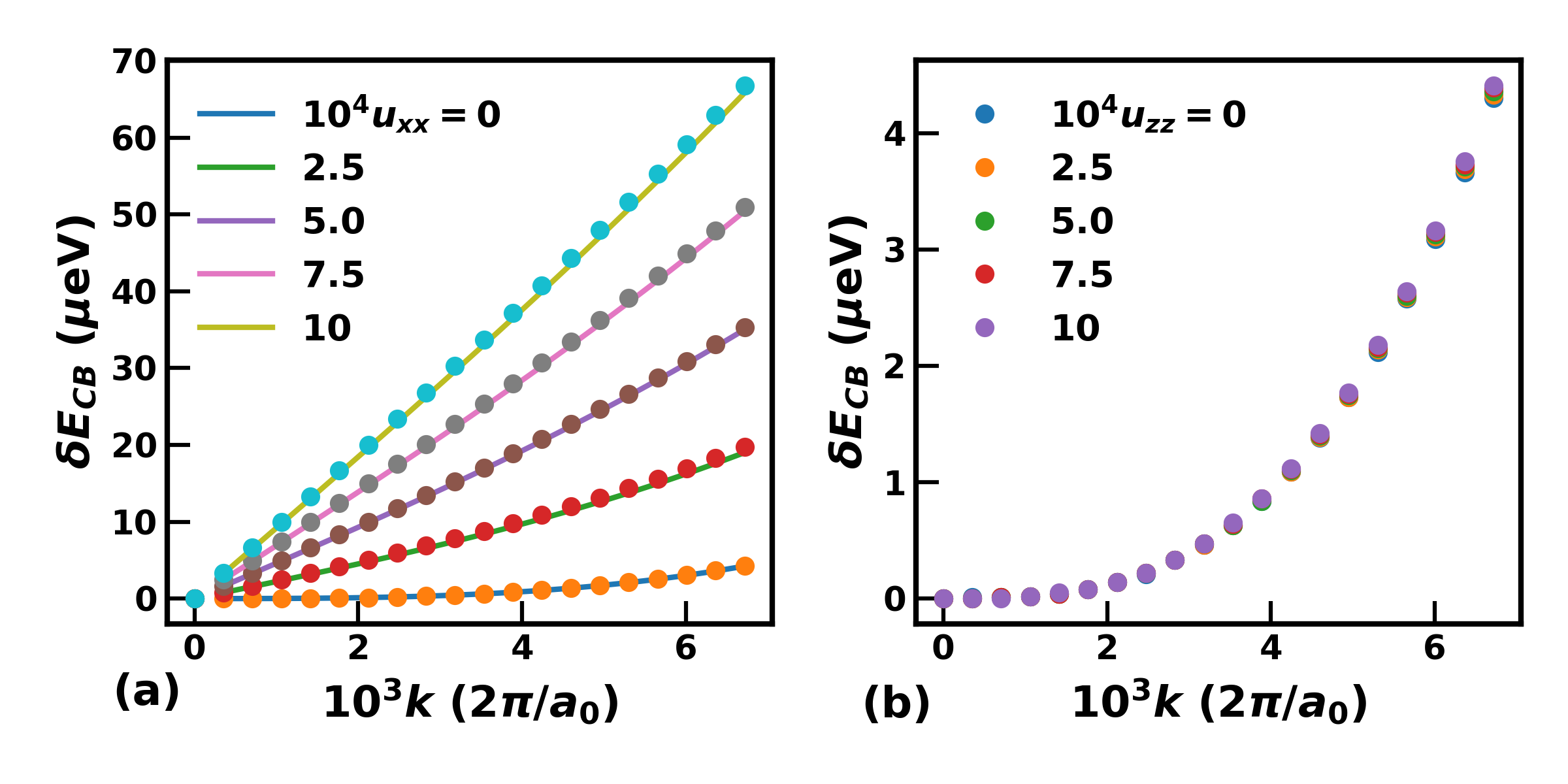}
\caption{Calculations of the conduction band spin splittings in GaAs under different strain components (a)
  $u_{xx}$ and (b) $u_{zz}$ (symbols). The lines are fits to   Eq.~\ref{dEeqS} (Dresselhaus and strain parts) used to determine the
  coefficients $|C_3|=1.65$~eVnm and $C_3'=0$. 
}
\label{uscan}
\end{figure}

We determined the strain coefficients $C_3$ and $C_3'$ by performing TB calculations for GaAs under strain and fitting to the expressions in Eq.~\ref{dEeqT}. $C_3'$ was calculated using a strain field with  $u_{xx}=0$, $u_{xz}=0$, and for different values for $u_{zz}$. The results are summarized in Fig.~\ref{uscan}(b). We find that $C_3'$ is essentially zero. 

In order to calculate $C_3$, we took $u_{zz}=0$, $u_{xz}=0$ and performed calculations for different  $u_{xx}$ [Fig.~\ref{uscan}(a)]. 
The TB calculations yield $|C_3|= 1.65$~eVnm. This  $C_3$ value agrees very well with the experimental one obtained from the QWR data in the main text. It is, however, twice as large as the one experimentally determined by Beck (0.81~eVnm),~\cite{Beck05a} and also larger  then the one estimated from D'yakonov's calculations.~\cite{Dyakonov86a,Pikus88a} 
\nPVS{I do not know the reasons for the discrepancy. I have to check if
everyone is using the appropriate strain (eng. vs. normal).}
Corresponding results for QW structures are presented in the main text.

%%%%%%%%%%%%%%%%%%%%%%%%%%%%%%%%%%%
% fitting strain coefficients

%%%%%%%%%%%%%%%%%%%%%%%%%%%%%%%%%%%

%%%%%%%%%%%%%%%%%%%%%%%%%%%%%%%%%
% SIFits
%%%%%%%%%%%%%%%%%%%%%%%%%%%%%%%%%

\section{Experimental Determination of strain-induced $C_3$ parameter}
\label{SIFits}
\subsection{QWs and QWRs data}

According to Eq.~\ref{dEeqT} (and to Eq.~2 of the main text), the  SO field for spins under the moving SAW field can be stated in the reference frame with axis $x=[110]$, $y=[\bar110]$ and $z=[001]$ as:

\begin{equation}
	\hbar\Omega_{SO}=\left[\gamma\left[\frac{\pi}{w_{z}}\right]^2-2F_zr_{41} - \frac{C_3u_{xx}}{2}\right]k_x.
	\label{eq:totalSO}
	\end{equation}

\noindent  Here, $k_x$ is given by Eq.~\ref{EqSIk}  and $w_{z} = d_{QW} + 2 d_{B}$ is the extension of the electron wave function along the $z$-direction given by the sum of the nominal QW width ($d_{QW}$) and the penetration depths in the upper and lower QW barriers ($d_{B}$). In the calculations that follow, we assume $d_{B}=2$~nm.  
The spin-orbit parameters are given by $\gamma$ = \SI{17e-30}{\electronvolt\metre\cubed} (Dresselhaus \cite{PVS158}), $r_{41}$ = -\SI{5.95e-20}{\elementarycharge\metre\squared} (Bychkov-Rashba \cite{PVS272}) and $C_3$ the strain parameter. 
 
\begin{figure}

	\includegraphics[width=1\columnwidth]{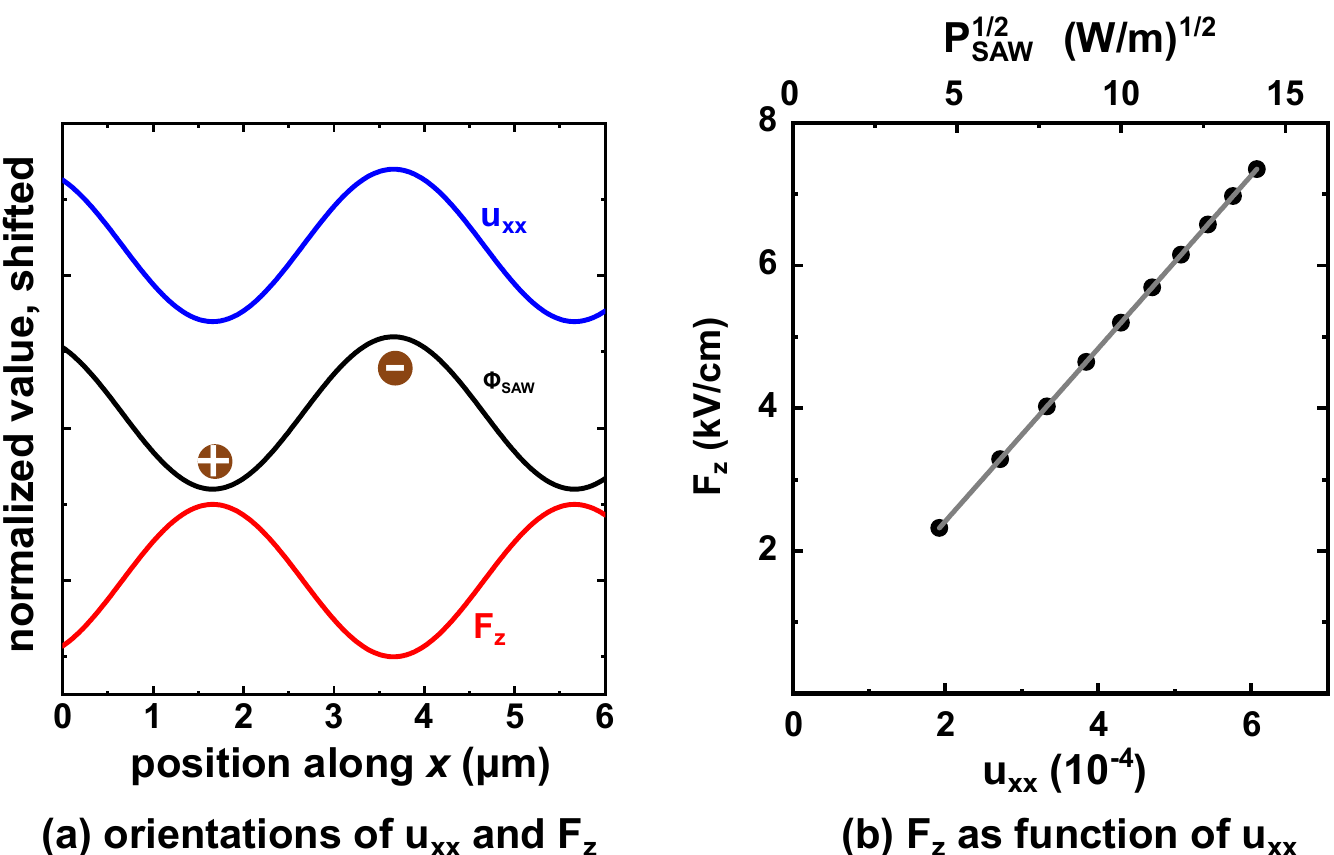}
	\caption{(a) Relative orientations of $\Phi_{SAW}$ (black), $F_z$ (red) and $u_{xx}$ (blue) along the QWR axis. The transport position of electrons (holes) in the case of ideal acoustic transport is depicted by brown circles with a minus (plus) sign. (b) Dependence of $\left|F_z\right|$ on $u_{xx}$. The upper horizontal scales displays the square root $\sqrt{P_\ell}\propto u_{xx,0}$, thus establishing the relationship between the strain amplitude and the linear acoustic power density generated by the IDT. }
	\label{fig:SI-SAWsimulations}
\end{figure}

\rPVS{
As mentioned in the main text, $F_z$ is proportional to $u_{xx}$. In order to determine the proportionality constant, }
{In order to construct the spin-orbit field as function of strain, the piezoelectric field can be expressed as a linear function of the strain $u_{xx}$. For this purpose,} 
we conducted numerical simulations of the acoustic and piezoelectric fields of  a SAW with a wavelength $\lSAW=4~\mu$m propagating along the \xdirnox-direction (see Ref.~\onlinecite{PVS156} for details). Figure~\ref{fig:SI-SAWsimulations}(a) shows the relative values of the SAW potential $\Phi_{SAW}$ (black), $u_{xx}$ (blue), and $F_z$ (red) at various positions/phases of the wave. The curves are normalized to their maximum values and shifted along the vertical axis for clarity. For this propagation direction,  the  maxima of $\FSAW$, corresponding to the transport phase of electrons (brown circles with minus sign), $F_z$ is negative (i.e., oriented along the $-z$ direction). The uniaxial strain $u_{xx}$ is positive at this position. Thus,  $u_{xx} > 0$ and $F_z < 0$ in Eq.~(\ref{eq:totalSO}). 

Figure~\ref{fig:SI-SAWsimulations}(b) shows the amplitude of the piezoelectric field $\left|F_z\right|$ as a function of the strain $u_{xx}$. The upper horizontal scale displays the corresponding values for  $\sqrt{P_\ell}\propto u_{xx}$, thus establishing the relationship between the strain amplitude and the linear acoustic power density $P_{\ell}$ excited by the IDT.  
The linear fit yields a slope of $r_\mathrm{SF}=\SI{12e3}{\kilo\volt\per\centi\metre}$. Note that the strain is small, so that the resulting field is on the order of \deleted[id=PH]{several} \SI{}{\kilo\volt\per\centi\metre}. Substitution of the simulation results modifies Eq.~(\ref{eq:totalSO}) in:

% --------------------------------------------------------------------------------
\begin{equation}
\hbar\Omega_{SO}=\left[\gamma\left[\frac{\pi}{w_{z}}\right]^2+
2r_\mathrm{SF} u_{xx} r_{41}-\frac{C_3u_{xx}}{2}\right]k_x.
\label{eq:totalSO_2}
\end{equation}
% --------------------------------------------------------------------------------

% --------------------------------------------------------------------------------
% --------------------------------------------------------------------------------

\noindent Note that the second term on the right-hand-side (corresponding to Bychkov-Rashba) turns positive due to the negative $F_z$.

The fits of the experimental data for QWRs in Fig.~4(b) of the main text  to Eq.~(\ref{fig:SI-SAWsimulations}) were carried out using an effective electron mass $m_e^*=0.067$ and  a SAW velocity $v_{SAW}$ = \SI{2904}{\metre\per\second} yielding $C_3$ = \SI{-2.6}{\electronvolt\nano\metre}. Table~\ref{table:Precession_QWR} summarizes the calculated values for the different components of the SO field for two different SAW powers. All SO contributions  are directed along the same direction (the  $+y$-axis). In addition,  the strain-related SO contribution is much larger than the one from the piezoelectric field.

\begin{table}
	\begin{tabular}{|p{2.5 cm}|c|c|c|c|c|}
		\hline
		&&&&&\\
		QW/QWR& $\Omega_D$ & $\Omega_{R}$   	& $\Omega_{S}$   & $\Omega_{exp}$  & $\Omega_{calc}$  \\         
		&  (GHz) &  (GHz)  	&  (GHz)  &  (GHz) &  (GHz) \\         
		\hline
		QWR, $\PSAW=7$~dBm      &     0.51      &     0.03      &  0.20 & 0.89 & 0.74 \\
		QWR, $\PSAW=21$~dBm      &     0.51      &     0.14      &  1.03 & 1.77 & 1.68\\
		QW, $\PSAW=7$~dBm &    2.20       &       0.03     & 0.20 &  1.66 & 2.43\\
		QW, $\PSAW=21$~dBm    &      2.20     &      0.14    & 1.03 & 1.71& 3.38  \\
		\hline
	\end{tabular}
	\caption{Spin-orbit precession frequencies in QWs and QWRs for different SAW powers.}
	\label{table:Precession_QWR}

\end{table}

% --------------------------------------------------------------------------------
\begin{figure}[htbp]
	%\centering

	\includegraphics[width=1\columnwidth]{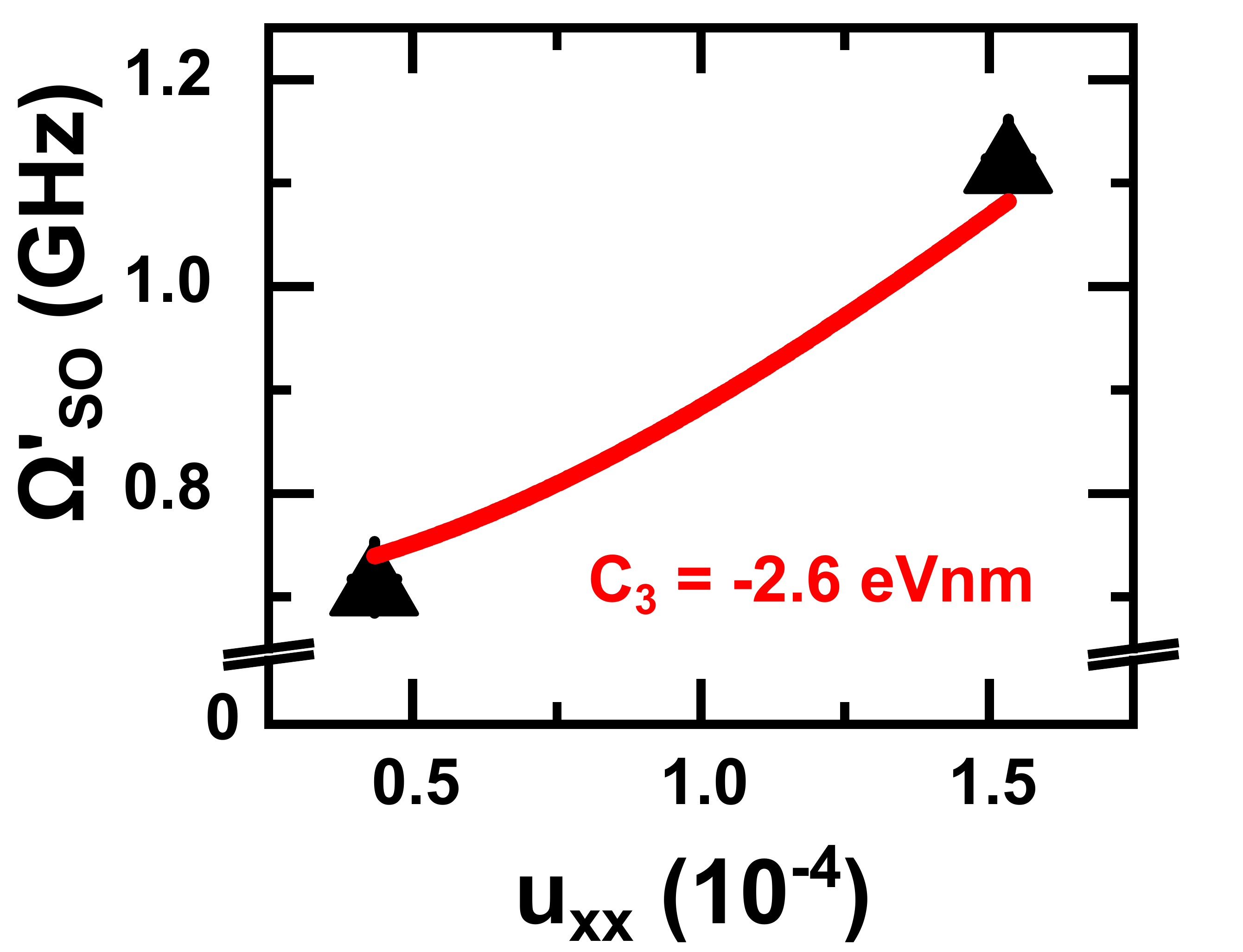}
	\caption{Fit of the DQD data to determine the $C_3$ parameter.}
	\label{fig:OmegaSO}
\end{figure}
% --------------------------------------------------------------------------------

\subsection{Calculation of $C_3$ parameter for DQDs}

Due to the different orientations of the Dresselhaus  and the SAW-related SO fields  expressed in Eq.~3 of the main text, the amplitude of the precession frequency experienced by spins in DQDs created by the superposition of SAW beams with the same intensity propagating along the $x$- and $y$-directions becomes:

\begin{equation}
\label{eq:SODQD}
\begin{split}
\left|\Omega^\prime_{SO}\right|=\sqrt{2\Omega_D^2+4\left[\Omega_{R}+\Omega_S\right]^2},
\end{split}
\end{equation}

\noindent where, as mentioned in the main paper, $\Omega_\mathrm{D}$, $\Omega_\mathrm{R}$, and $\Omega_\mathrm{S}$ are the magnitudes of the precession frequency for a single SAW beam along the $x$-direction. As in the previous case, the amplitude of the transverse piezoelectric field can be related to the strain amplitude $u_{xx}$ by a factor $r_\mathrm{SF}=21\times 10^3$~kV/cm. \deleted[id=PH]{\cPVS{Paul, James: Important!!! why is this factor much larger for the DQD than for the QWR?}}
Due to the quadratic dependence of $\Omega^\prime_{SO}$ on $\Omega_\mathrm{D}$ in Eq.~(\ref{eq:SODQD}), it is not possible to determine from the DQD data in Fig.~4(b) of the main paper the relative signs of the Dresselhaus and SAW-related SO fields. The red line in Fig.~\ref{fig:OmegaSO} shows, however, that the SAW power dependence of $\Omega^\prime_{SO}$ is reasonably well accounted for using a value of $C_3$ = \SI{-2.6}{\electronvolt\nano\metre} that is comprable to the QWR results.  Table~\ref{table:Precession_DQD} summarizes the values of the different SO-precession frequencies for DQDs.

\begin{table}
	\begin{tabular}{|p{2.5cm}|c|c|c|c|c|}
		\hline
		&&&&\\
		DQDs & $\Omega_D^\prime$ & $\Omega_{R}^\prime$   	& $\Omega_{S}^\prime$   & $\Omega_{exp}^\prime$  & $\Omega_{calc}^\prime$ \\         
		&  (GHz) &  (GHz)  	& (GHz)  & (GHz) &  (GHz)\\         
%		\hline

%		\hline
		\hline
		$\PSAW=2$~W/m ($C_3=-2.0$)      &     0.53      &     -0.05      &  -0.23 & 0.71 & 0.60 \\
		$\PSAW=25$~W/m  ($C_3=-2.0$)     &     0.53      &     -0.20      & -0.79 & 1.11 & 1.12\\
		\hline
	\end{tabular}
	\caption{Spin-orbit precession frequencies determined for transport by DQDs under different SAW powers.}
	\label{table:Precession_DQD}
\end{table}

%%%%%%%%%%%%%%%%%%%%%%%%%%%%%%%%%
% SIMonteCarlo
%%%%%%%%%%%%%%%%%%%%%%%%%%%%%%%%%
\section{Monte-Carlo simulations of the acoustic spin transport}
\label{SIMonteCarlo}
 
We study, in this section, the spin dynamics during acoustic transport in narrow channels using Monte-Carlo simulations of the spin trajectories. Similar Monte-Carlo investigations of the confinement effects on the spin dynamics were carried out for diffusing spins by Kiselev and Kim
\cite{Kiselev_PRB61_13115_00,Kiselev_pssb221_491_00}. 
In contrast to the results presented here, those previous studies were restricted to the transport dynamics for short spin coherence times, where spatial precession oscillations are not observed. In addition, these studies did not address the impact of carrier drift under an external field on the spin dynamics. 

The Monte-Carlo simulations  were performed by assuming that $z|[001]$-oriented electron spins are optically excited at the position $(x,y)=(0,0)$ at the center of a channel with width $w_y$. A  surface acoustic wave (SAW) propagating with phase velocity $\vSAW$ along $x||[110]$ collects the spins and transports them along the $x$-direction with the SAW velocity. During transport, the electron spins precess around the SO-field with angular precession frequency $\OSO$ given by:

\begin{equation} 
\Omega_\mathrm{SO}= (1-|r_s|)\Omega_\mathrm{BIA} + r_s \Omega_\mathrm{SIA},
\label{EqSMOSO}
\end{equation} 

\noindent where   $\Omega_\mathrm{BIA}$ and $\Omega_\mathrm{SIA}$ are the components of the SO-field due to the bulk (BIA) and structural (SIA) inversion asymmetry, respectively, and $r_s$ is a factor weighting the amplitude of the two components (see further details below). For small magnitudes of the electron momentum $\hbar \bf k$,  these components can be expressed in the ${\bf k}=(k_x, k_y)=k(\cos\theta, \sin\theta)$  basis, where $\theta$ is the propagation angle with respect to the $x$ axis,  as: 

\begin{eqnarray}
	   \Omega_\mathrm{BIA}&=& \underbrace{c_\mathrm{BIA}k}_{\Omega_\mathrm{BIA}^\mathrm{(max)}} \begin{bmatrix}
                +\sin(\theta)\\
                +\cos(\theta)
                \end{bmatrix}     
                \label{FigSI-OBIA}  \\ 
	\Omega_\mathrm{SIA}&=& \underbrace{c_\mathrm{SIA}k}_{\Omega_\mathrm{SIA}^\mathrm{(max)}} \begin{bmatrix}
                -\sin(\theta)\\
                +\cos(\theta)
                \end{bmatrix}   
         \label{FigSI-OSIA}
                \end{eqnarray}
    
\noindent  Here, $c_\mathrm{i}$ ($i=\mathrm{BIA}, \mathrm{SIA}$) are material constants associated with the BIA and SIA components determining the precession frequencies $\Omega_\mathrm{i}^\mathrm{(max)}=c_i k$ (i.e., only linear terms in $k$ are taken into account). 

Due to the dependence on carrier  momentum, the determination of the spin dynamics under the SO fields requires the knowledge of the electron trajectory during the motion. The latter was determined based on the following assumptions:

% ---------------------------------------------
% Figure SMFigL
\begin{figure}[tbhp]
\includegraphics[width=.9\columnwidth]{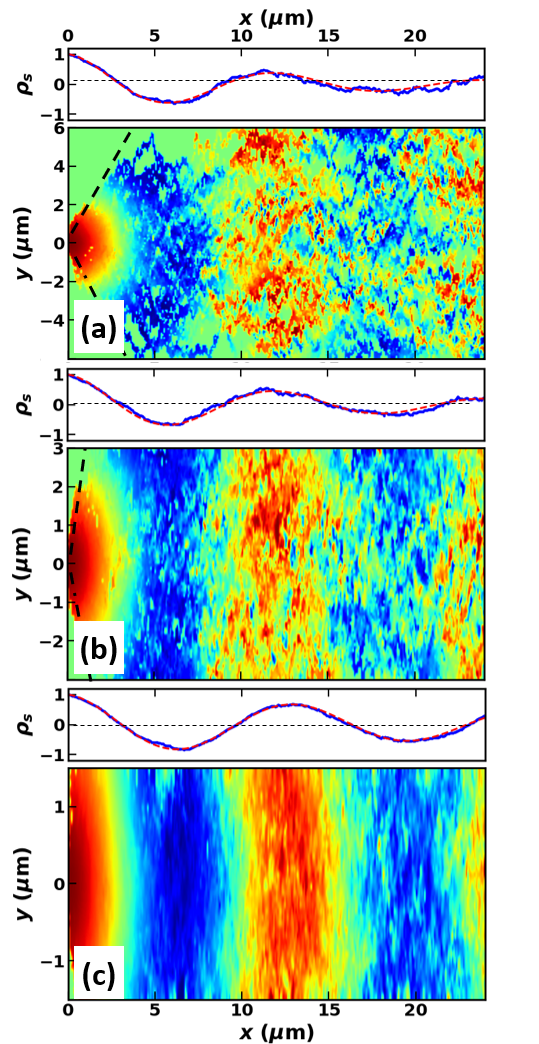}
\caption{
Spin polarization profiles during acoustic transport in channels with width  (a) $w_y=L^{(2D)}_{SO}=12~\mu$m, (b) $w_y= 6~\mu$m,  and (c) $w_y=3~\mu$m. The profiles  result from Monte-Carlo simulations of the spin   transport under a spin-orbit field with SIA asymmetry quantified by an effective  spin-orbit length $L_\mathrm{SO}=12~\mu$m. The simulations assume a temperature of 10~K and carrier mobility of $4~m^2/(Vs)$.  The colors encode the spin projection along $z$. The upper plots display the spin polarization $\rho_s$ obtained by integrating the color maps within the range $|y|<2~\mu$m.
}
\label{SIFigSpin4}
\end{figure}
% ---------------------------------------------
% ---------------------------------------------
% FigureSIFigSpin2
\begin{figure*}%[tbhp]
\includegraphics[width=.85\textwidth]{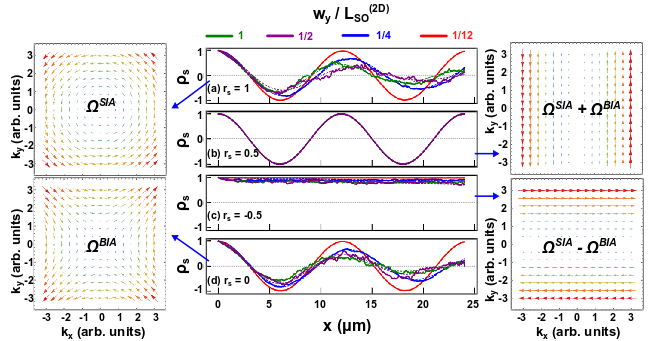}
\caption{
Spin polarization profiles along the SAW transport path calculated for different ratios $w_y/L_{SO}$  and 
(a) $r_s=1$ [$\Omega_\mathrm{SO}=\Omega_\mathrm{SIA}$, cf. Eq.~(\ref{FigSI-OSIA})]  
(b) $r_s=0.5 $ ($\Omega_\mathrm{SO}=\left(\Omega_\mathrm{SIA}+\Omega_\mathrm{SIA}\right)/2$),
(c) $r_s=-0.5 $ ($\Omega_\mathrm{SO}=\left(\Omega_\mathrm{SIA}-\Omega_\mathrm{SIA}\right)/2$), and
(d) $r_s=0$ [$\Omega_\mathrm{SO}=\Omega_\mathrm{BIA}$, cf. Eq.~(\ref{FigSI-OBIA})].
(b) and (c) are the spin helix modes with ${\bf \Omega_\mathrm{SO}}$ oriented along the $y$- and $x$-directions, respectively.
 The right and left inset display the corresponding momentum dependence of the spin-orbit field for these four configurations.
}
\label{SIFigSpin5}
\end{figure*}
% ---------------------------------------------

%\end{document}

\begin{itemize}
\item the propagation velocity $\vSAW$ along $x$ imparts  a momentum along this direction given by $\hbar k_x = m_e^* \vSAW$, where $m_e^*$ is the electron effective mass;

\item the spins are free to diffuse along $y||[\bar1 1 0]$ (i.e., along the SAW wave fronts) with a velocity $v_y$ (and corresponding momentum $ m_e^*v_y$)  determined from a thermal distribution given by:

\begin{equation}
f(v_y) \delta v_y = \left[ \frac{m_e^*}{2\pi k_B T} \right]^{1/2}  e^{-{\frac {m_e^* v_y^{2}}{2 k_B T}}} =
\frac{1}{\sqrt{\pi}v_p} e^{-v^2/v_p^2} \delta v_y ,
\end{equation}

\noindent where  $v_p = \sqrt{\frac{2 k_B T}{m_e^*}}$ and $T$ is the temperature. The diffusion along $y$ proceeds until the spins are back-reflected at the edges of the channel.

\item within the channel boundaries, the electron spins precess around the spin orbit field with an angular precession frequency $\OSO$ [cf. Eq.~(\ref{EqSMOSO})] determined by the momentum $\hbar \bf k$ until they are scattered with a  scattering time $\tau$.  The scattering time $\tau$ is determined from the mobility $\mu$ according to:
	$\tau = m_e^* \mu/q_e$, where $q_e$ is the electron charge. The motion along $y$ is  randomized at each scattering event.
	
\item Due to the linear dependence of the precession frequencies on $\bf k$, the spin precession angle for propagation along a fixed direction only depends on the distance between the initial and final points of the trajectory.  
If $(\cos{\theta}, \sin{\theta})\Delta \ell$ is the displacement vector between two successive scattering events,
than the corresponding spin precession angle under a SO field $\Omega_\mathrm{SIA}$ will be  $\delta \phi_s =  2\pi \Delta \ell /L^\mathrm{SIA}_{SO}$. Here,  $L^\mathrm{SIA}_{SO} = m_e^* c^\mathrm{SIA}/\hbar$ is the SO length, which only depends on material properties. These results apply for  $\Delta\ell<<L^{SIA}_\mathrm{LO}$ and are also valid for the BIA component. The spin  dynamics results to be shown below were calculated  assuming  $L^{SIA}_\mathrm{SO}=L^{BIA}_\mathrm{LO}=L^{(2D)}_\mathrm{SO}=12~\mu$m.

\end{itemize}

% ---------------------------------------------

\begin{figure}[tbhp]

\includegraphics[width=\columnwidth]{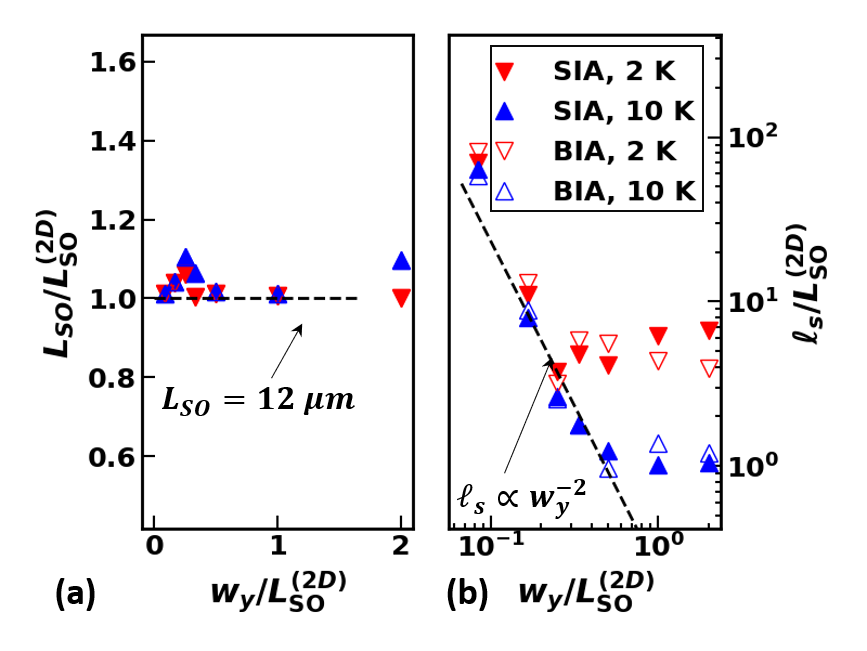}
\caption{
Dependence of the fitted (a) spin-orbit precession ratio  $\LSO/\LSO^{(2D)}$ and (b) spin dephasing length ($\ell_s$) normalized to the precession period $\LSO^{(2D)}=2 \pi\vSAW \OSO$. The parameters were obtained from fits to  Monte-Carlo simulations of the acoustic spin transport along channels with different widths $w$. Results are shown for transport at 2~K and 10~K under a BIA-type and SIA-type spin-orbit field. The $\LSO$  and $\ell_{s}$ values were obtained by fitting the calculated spin polarization profiles  to Eq.~(1) of the main paper.}
\label{SIFigSpin6}
\end{figure}
% ---------------------------------------------

Figures~\ref{SIFigSpin4}(a)-(c) compare spin polarization maps calculated for acoustic transport under a SO field with SIA symmetry ($\Omega_{SIA}$) along channels with different widths $w_y$.  In each map,  the colors encode the $z$ projection of the spin vector calculated by averaging over  40 Monte-Carlo realizations. In the wide channels [Figs.~\ref{SIFigSpin4}(a)], the individual spin trajectories can still be identified. Here, one sees that  the spins initially propagate radially from the excitation spot $(x, y)=(0,0)$ within a cone (dashed lines) defined by the ratio between the average thermal velocity and the SAW velocity. The spin  polarization within this cone oscillates along the radial direction with a precession period given by $L_{SO}$. For longer transport distance along $x$, a increasing number of trajectories with different spin polarization superimpose, leading to a decay of the average spin polarization.

The panels on top of the maps in Fig.~\ref{SIFigSpin4} display the average spin polarization $\rho_s$ obtained by integrating the spin projection maps for $|y|<2~\mu$m. The superimposed dashed lines are fits to Eq.~(1) of the main text, which yield the effective SO length $L_{SO}=\Omega_{SO}/\vSAW$ as well as spin decay length $\ell_s$ (see below). For the widest channel in Fig.~\ref{SIFigSpin4}, the spin polarization decays with a distance $\ell_s$ corresponding roughly to $\LSO^\mathrm{(2D)}$. However, $\ell_s$ increases significantly with decreasing channel width, as shown in the upper plots of  Figs.~\ref{SIFigSpin4}(b) and \ref{SIFigSpin4}(c).

One interesting question is how the spin dynamics depends on the symmetry of the SO interaction. In order to address this question, Fig.~\ref{SIFigSpin5} displays  spin polarization profiles calculated for different ratios $r_s$ between the SIA and BIA SO contributions. The left and right panels illustrate the momentum dependence of the SO coupling for each of the cases.  
The calculations were carried out for different channel widths $w_y$.
Figures~\ref{SIFigSpin5}(a) and \ref{SIFigSpin5}(d) are for the pure BIA and SIA contributions, respectively. Within the calculation accuracy, one observes a similar behavior for the two types of asymmetry with  the  spin decay length $\ell_s$ increasing with decreasing channel width. Figures~\ref{SIFigSpin5}(b) and \ref{SIFigSpin5}(c) show the  simulation results for $r_s=0.5$ and $r_s=-0.5$, which correspond to the spin helix modes with ${\bf \Omega_\mathrm{SO}}$ parallel to $y$ and $x$, respectively (cf.~right panels) \cite{Bernevig_PRL97_236601_06,Koralek_N458_610_09,Walser2012}. In both cases, the spin decay reduces significantly.

The dependence of the  spin precession period ($\LSO$) and decay lengths obtained from fits of the calculated profiles to Eq.~(1) of the main text are summarized in Figs.~\ref{SIFigSpin6}(a) and (b), respectively. Results are presented for simulations under  SIA (filled symbols) and BIA (open symbols) SO fields at the temperatures of 2~K and 10~K, which yield different carrier mean-free-path $\ell_p$. The fitted precession periods are found to coincide within 10\% with the period  $\LSO^{(2D)}=12~\mu$m quantifying the SO interaction strength in the simulations.  Such a  behavior is indeed expected from Eqs.~(\ref{EqSMOSO}) for $r_s>0$, when the magnitude $|\OSO|$ is constant. Since the time $\Delta t$  for propagation over a short distance $\Delta r=(\Delta x, \Delta y)$,  given by $\Delta t = \Delta r/v = m_e^* \Delta r/ (\hbar k)$, is inversely proportional to $k$, then the  precession angle around $y$ can be expressed as    $\delta \phi = \OSO_{y} \Delta t=\OSO\cos\theta \Delta t =  (\hbar/m_e^*) \OSO \Delta x$.
As a consequence, the precession period along $x$ does not depend on the spin propagation  angle $\theta$.  

The dependence of $\ell_s$ on $w_y$  displayed in Fig.~\ref{SIFigSpin6}(b) shows two regions with different behaviors. For wide channels, $\ell_s$ saturates at a value
\begin{equation}
\ell_s \propto \vSAW * \frac{(\LSO^{(2D)})^2}{  2 \pi v_y^2 \tau_p },
\end{equation}

\noindent which reduces with increasing temperature. For narrow channels, in contrasts, $\ell_s$ is essentially  independent of temperature and increases according to $\ell_s\propto w^{-2}$. Such a dependence agrees with previous simulation results from 
Refs.~\onlinecite{Kiselev_PRB61_13115_00,Kiselev_pssb221_491_00} 
and is attributed to the increased role of motional narrowing for $w_y<\LSO^{(2D)}$.

The results of this section thus show that while the spin precession frequency during acoustic transport is essentially independent of the channel width, the spin decoherence lengths increase significantly in narrow channels ($w_y<<L_\mathrm{SO}$).

%\end{document}

\dPVS{
% --------------------------------------------------------------------------------
\begin{figure*}
	\includegraphics[width=.65\textwidth]{Transport}
	\caption{
	Maps of PL emission excited at $(x,y)=(0,0)$ under a SAW with $P_{SAW}$ = \dbm{10} and diferent separations $\Delta x$ between the laser excitation spot  and detection position on a trap for the (a) QWR,  $\Delta x=9~\mu$m, (b) QWR,  $\Delta x=20~\mu$m, (c) QW, and (d) QW,  $\Delta x=20~\mu$m.  
	(e) Typical time-resolved PL emission from a remote trap in the QWR path in the absence of a SAW (black) and under SAW powers of $P_\mathrm{SAW}=\dbm{15}$ (red) and \dbm{20}. (f) Corresponding time-resolved profiles for a QW in the absence of a SAW (black) and under SAW powers of of \dbm{12} (red) and \dbm{14} (blue).  
	 {(\bf \color{red}{Paul: what is the transport distance in the time resolved spectra?)}}
	}
	\label{fig:SI-Transport}
\end{figure*}
% --------------------------------------------------------------------------------

%%%%%%%%%%%%%%%%%%%%%%%%%%%%%%%%%
% SITransport
%%%%%%%%%%%%%%%%%%%%%%%%%%%%%%%%%
\section{Transport efficiency}
\label{SITransport}

In order to compare the carrier transport efficiencies in  QWRs and  QWs, Figs.~\ref{fig:SI-Transport}(a)-(d) depict mappings of the PL emission under a SAW of power \dbm{10}. In a typical measurement and for relative long transport distances, the QWR emission is observed only at the excitation spot and at a remote location corresponding to a trapping center [cf.~Fig.~\ref{fig:SI-Transport}(b)]. Note that the lateral extent of the PL cloud is determined by the spatial resolution of the detection setup. Moving the excitation spot closer to the excitation spot [cf.~Fig.~\ref{fig:SI-Transport}(a)] results in an overlap of the emission clouds at the two positions. 
A similar condition as the latter is typically observed for long distances as well in the QW in Fig.~\ref{fig:SI-Transport}(c) and \ref{fig:SI-Transport}(d). In the QW, a tail of emission is observed from excitation spot to the position of the remote PL, thus indicating that a larger number of trapping centers are present in the QW. Note that the smaller thickness of the QW with respect to the QWR makes the carriers in the QW more susceptible to potential fluctuations due to the monolayer fluctuations, which can possibly explain the observed difference.

In order to attest the efficiency of carrier storage in the SAW potential, Figs.~\ref{fig:SI-Transport}(e) and \ref{fig:SI-Transport}(f) show time-resolved PL measurements detected on a typical trap in the QWR and the QW, respectively. In both figures, the black curve shows that without SAW, only a small background is observed, thereby confirming that the remote emission originates from carriers transport by the SAW. Applying a SAW along the QWR [\ref{fig:SI-Transport}(e)], an intermediate SAW power of \dbm{15} (red) results in oscillations of the emission. The ratio between baseline and amplitude is a measure of the transport efficiency and depends on the SAW power (not shown here). For high acoustic powers (blue line), the phase of the trap emission changes, indicating that the trapping nature now preferential traps carrier of the opposite polarity. In this condition, the ratio between baseline and amplitude is very large with respect to the baseline, meaning that remote emission now takes place continuous during all SAW phases, which indicates a poor transport efficiency. 

A similar case is observed in the QW [\ref{fig:SI-Transport}(f)]. Here, for intermediate SAW powers (red line) the remote QW emission oscillates, where a baseline is observed as well. Increasing the SAW power largely reduces the emission (blue line), while it mostly results in a flat line of the emission. This indicates again that emission takes place in all SAW phases and therefore that carriers are poorly stored in the SAW potential.

The results shown in this section indicate that the efficiency of the SAW potential in both the QW and the QWR depends on the experimental conditions. However, no clear improvement of the transport efficiency in either the QW or the QWR with respect to the other is observed. The absence of a tail of emission in the QWR between excitation and remote emission for longer transport distances indicates, however, a larger transport efficiency in the QWR due to the suppression of trapping centers.

}

%\printbibliography
%\bibliography{mypapers,literature}
%\bibliography{mypapers,literature}
%\bibliography{word}
%%%%%%%%%%%%%%%%%%%%%%%
\IfFileExists{literature.bib}
{    \def\litdir{.}}
{   
     \IfFileExists{c:/myfiles/jabref/literature.bib}
    {               \def\litdir{c:/myfiles/jabref}      }
    {	         \def\litdir{x:/sawoptik_databases/jabref}     }
}